\documentstyle[prl,eqsecnum,aps,epsf,floats]{revtex}
\begin{document}

\title{Comprehensive study of Frohlich polaron}

\author{A.S.~Mishchenko$^{1,2}$, N.V.~Prokof'ev$^{1,3}$, A.~Sakamoto$^2$, 
and B.V.~Svistunov$^1$}

\address{$^1$ Russian Research Center ``Kurchatov Institute", 123182 Moscow, 
Russia \\
$^2$ Department of Applied Physics, The University of Tokyo,
7-3-1 Hongo, Bunkyo-ku, Tokyo 113, Japan \\
$^3$ Department of Physics, University of Massachusetts, Amherst, MA 01003, USA}

\date{\today}

\maketitle

\begin{abstract}
A detailed study of the Fr\"{o}hlich polaron model is performed on the
basis of diagrammatic quantum Monte Carlo method \cite{PS}.
The method is further developed both quantitatively (performance)
and qualitatively (new estimators), and is enhanced by spectral
analysis of the polaron Green function, within a novel approach.  
We present the best up to date results for the binding energy, 
and for the first time make available precise data for the effective mass, 
including the region of intermediate and strong couplings. We look at the 
structure of the polaron cloud and answer such questions as the average 
number of phonons in the cloud and their number/momentum distribution.
The spectral analysis reveals non-trivial structure of the spectral
density at intermediate and large coupling: the spectral continuum
features pronounced peaks that we attribute to unstable excited
states of the polaron.
\end{abstract}

\vskip2pc

\section{Introduction}
\label{sec:1}

The polaron problem (for an introduction see Ref.~\onlinecite{polaron})
has originally emerged in the solid state physics as a problem of 
electron moving in a (dielectric) medium. It became clear, however, 
that this problem is of essential general-physical interest, as a model 
of a quantum object strongly coupled to an environment. 
Starting from the work of Landau \cite{Landau33}, the polaron problem
has been attracting a permanent attention, serving as a testing
ground of new non-perturbative methods. 

The most popular model in the polaron problem is the so-called
Fr\"{o}hlich Hamiltonian describing an electron
coupled to non-dispersive (optical) phonons of a dielectric
medium via its polarization (Plank's constant and 
electron mass are set equal to unity) \cite{Frohlich}:
\begin{equation}
H \, = \, H_{\mbox{\scriptsize e}} +
H_{\mbox{\scriptsize ph}} +H_{\mbox{\scriptsize e-ph}} \; ,
\label{H}
\end{equation}
\begin{equation}
H_{\mbox{\scriptsize e}} \, = \, 
\sum_{\bf k} \, \frac{k^2}{2} \, a^{\dag}_{\bf k} a^{ }_{\bf k} \; , 
\label{e}
\end{equation}
\begin{equation}
H_{\mbox{\scriptsize ph}} \, = \, 
\sum_{\bf q} \, \omega_q \, b^{\dag}_{\bf q}  b^{ }_{\bf q} \; ,
\label{ph}
\end{equation}
\begin{equation}
H_{\mbox{\scriptsize e-ph}} \, = \, 
\sum_{{\bf k},{\bf q}} \, V({\bf q}) \, 
\left( b^{\dag}_{\bf q} - b^{ }_{-{\bf q}}   \right) \,
a^{\dag}_{{\bf k}-{\bf q}} a^{ }_{\bf k} \; ,
\label{e-ph}
\end{equation}
\begin{equation}
V({\bf q}) \, = \, i \, \left( 2 \sqrt{2} \alpha \pi \right)^{1/2} \,  
\frac{1}{q} \; .
\label{V}
\end{equation}
In Eqs.~(\ref{H}-\ref{V}), $a^{ }_{\bf k}$ and $b^{ }_{\bf q}$ are the 
annihilation 
operators for the electron with momentum $\bf k$ and for the phonon 
with momentum $\bf q$, respectively, $\omega_q \equiv \omega_0$ is 
the $\bf q$-independent phonon frequency, which can be set equal to 
unity without loss of generality, $\alpha$ is a dimensionless coupling 
constant.

Despite a lot of work addressed to the Fr\"{o}hlich Hamiltonian,
the model is still far from being completely understood. 
In the most interesting region - at intermediate and large
values of $\alpha$, almost all available treatments are of
variational character. Hence, these treatments, even if consistent 
with each other, can not guarantee quantitative and 
qualitative reliability of the results. Moreover, some
treatments are known to be in qualitative disagreement with
the others. As a characteristic example, note that certain
approaches suggest that the polaron states at small and
large $\alpha$'s are of qualitatively different nature, and
there should occur a sort of phase transition in the
parameter $\alpha$ 
\cite{Gross,Matz,Luttinger,Manka,Manka80,Lepine,Feranchuk} 
(see, however, discussion in Ref.~\onlinecite{Peeters}). 
The study of such important issues as electron $Z$-factor and the
structure of the polaronic cloud was also restricted to the
perturbation theory and variational treatments at small momenta
\cite{Feynman_book,Feynman55,Schultz59,LLP53}. 
It remained unclear what are the limits of applicability 
of these results, and whether they correctly describe the physics of 
polarons in the most important range of intermediate $\alpha$.

Recently, a method of diagrammatic quantum Monte Carlo (MC)
was developed, which is very efficient for the polaron-like
problems \cite{PS}. The method allows direct simulation
of entities specified in terms of (positive definite) diagrammatic
expansions. In Ref.~\onlinecite{PS} the polaron Green function was 
simulated, and the results were used to extract the polaron
spectrum.

In the present paper, we employ the diagrammatic Monte Carlo
scheme of Ref.~\onlinecite{PS} for a detailed study of the Fr\"{o}hlich 
model. We significantly enhanced the original scheme by (i) introducing
$N$-phonon Green functions (with $2N$ external phonon lines), which 
are simulated in one and the same MC process with the ordinary 
($0$-phonon) Green function, (ii) developing a powerful procedure
of spectral analysis of the Green function. The $N$-phonon Green 
functions allow us to consider the structure of the phonon cloud
and facilitate obtaining polaron parameters at large $\alpha$,
where the polaron is essentially a many-phonon object.
In particular, direct estimators for the energy, effective mass,
group velocity, and $Z$-factors can be constructed.
The spectral analysis of the Green function gives the 
most complete information about the polaron, including the
possibility to reveal stable and metastable excited states, 
if any.

The paper is organized as follows. In Section \ref{sec:2}
we introduce the set of Green functions, describe the
corresponding diagrammatic series, and discuss how they are 
related to the polaron parameters. In Section \ref{sec:3} 
we describe qualitatively the Monte Carlo procedure [the
quantitative discussion of the updates is given in the Appendix A].
The calculated properties of the polaron (energy, effective
mass, structure of the polaronic cloud, ets.) are presented in Section
\ref{sec:4}. In Section \ref{sec:5} we analyze excited states 
of the polaron by restoring the spectral density of the Green 
function by a novel method of the spectral analysis [detailed 
description of the method is presented in Appendix B].

\section{Green functions and diagrams}
\label{sec:2}

In this section we introduce basic entities and establish their 
relations, which will be utilized in the rest of the paper.

We start with the standard Green function of the polaron in
the momentum (${\bf k}$) -- imaginary-time ($\tau$) representation:
\begin{equation}
G({\bf k}, \tau) \, = \, \langle \mbox{vac} \vert \,
a^{ }_{\bf k}(\tau ) a^{\dag}_{\bf k}(0) \,
\vert \mbox{vac} \rangle \; , \; \; \tau \geq 0 \; ,
\label{G}
\end{equation}

\begin{equation}
a^{ }_{\bf k}(\tau ) \, = \, e^{H \tau} a^{ }_{\bf k} e^{-H \tau}  \; .
\label{a}
\end{equation}
Here $\vert \mbox{vac} \rangle$ is the vacuum state.

The physical information that $G({\bf k}, \tau)$ contains is
clear from the expansion
\begin{equation}
G({\bf k}, \tau) \, = \, \sum_{\nu} \,
\vert \langle \nu \vert a^{\dag}_{\bf k} \vert \mbox{vac} \rangle \vert^2 
\, e^{-\left( E_{\nu}({\bf k}) - E_0 \right) \tau}\; ,
\label{L}
\end{equation}
where $\{ \vert \nu \rangle \}$ is a complete set of eigenstates of the
Hamiltonian $H$ in the sector of given ${\bf k}$, i.e. 
~$H \, \vert \nu ({\bf k}) \rangle = 
E_{\nu}({\bf k}) \, \vert \nu ({\bf k}) \rangle$, 
~$H \, \vert \mbox{vac} \rangle = 
E_0 \, \vert \mbox{vac} \rangle$. Since in our model $E_0 = 0$, we omit
it below.  Rewriting 
Eq.~(\ref{L}) as
\begin{equation}
G({\bf k}, \tau) \, = \, \int_0^{\infty} d \omega \, g_{\bf k} (\omega) \,
e^{- \omega \tau} \; ,
\label{Fr}
\end{equation}
\begin{equation}
g_{\bf k} (\omega) \, = \, \sum_{\nu} \,
\delta(\omega - E_{\nu}({\bf k})) \;
\vert \langle \nu \vert a^{\dag}_{\bf k} \vert \mbox{vac} \rangle \vert^2 
\; ,
\label{g}
\end{equation}
one defines the spectral function $g_{\bf k} (\omega)$ which
has poles (sharp peaks) at frequencies corresponding to stable
(metastable) particle-like states. Hence, if at a given ${\bf k}$
there exists a stable polaron with the energy $E({\bf k})$,
the spectral function reads
\begin{equation}
g_{\bf k} (\omega) \, = \, Z^{({\bf k})} \,
\delta(\omega - E({\bf k})) \, + \, \ldots \; ,
\label{g1}
\end{equation}
where
\begin{equation}
Z^{({\bf k} )} \, = \, \vert \, \langle \mbox{~polaron~}({\bf k})~ 
\vert \mbox{~free electron~}({\bf k})~
\rangle \, \vert^2 \; .
\label{Z}
\end{equation}
Moreover, if the polaron state is the ground state, its energy and 
$Z$-factor are ``projected out" by the Green function 
behavior at long times:
\begin{equation}
G({\bf k},\tau \gg \omega_0^{-1} ) \; \to \; 
Z^{({\bf k} )} \, e^{ -E({\bf k}) \tau } \; .
\label{asym}
\end{equation}

Along with the standard polaron Green function (\ref{G}), it is reasonable
to introduce the $N$-phonon Green function
\begin{equation}
G_N({\bf k}, \tau ; \, {\bf q}_1, \ldots , {\bf q}_N ) \, = \, 
\langle \mbox{vac} \vert \,
b^{ }_{{\bf q}_N}({\tau}) \cdots b^{ }_{{\bf q}_1}({\tau})\,
a^{ }_{\bf p}(\tau ) a^{\dag}_{\bf p}(0) \,
b^{\dag}_{{\bf q}_1}(0) \cdots b^{\dag}_{{\bf q}_N}(0)
\vert \mbox{vac} \rangle \; , \; \; 
{\bf p} \, = \, {\bf k} - \sum_{j=1}^N {\bf q}_j \; .
\label{G_N}
\end{equation}
Relations (\ref{L}-\ref{asym}) are readily generalized to the
case of $N$-phonon Green function. In particular, the $N$-phonon 
$Z$-factor for the stable (groundstate) polaron with momentum
${\bf k}$,
\begin{equation}
Z^{({\bf k} )}_{N} ({\bf q}_1, \ldots , {\bf q}_N ) \, = 
\, \vert \, \langle \mbox{~polaron~}({\bf k})~
\vert \mbox{~free electron~}({\bf p}) \, + \, 
\mbox{free phonons~}({\bf q}_1, \ldots , {\bf q}_N )~
\rangle \, \vert^2 \; 
\label{ZN}
\end{equation}
[the momentum ${\bf p}$ is defined as in Eq.~(\ref{G_N})], is given by
\begin{equation}
G_N({\bf k},\tau \gg \omega_0^{-1} ; 
\, {\bf q}_1, \ldots , {\bf q}_N ) \; \to \; 
Z^{({\bf k} )}_{N} ({\bf q}_1, \ldots , {\bf q}_N ) \; 
e^{ -E({\bf k}) \tau } \; .
\label{asym_N}
\end{equation}

Our MC procedure of simulating Green functions will utilize a
standard diagrammatic expansion - Matsubara technique at $T=0$.
The diagrams (see Figs.~\ref{fig:fig1}, \ref{fig:fig2}) 
are built of the following elements:
\begin{figure}
\epsfxsize=0.75\textwidth
\epsfbox{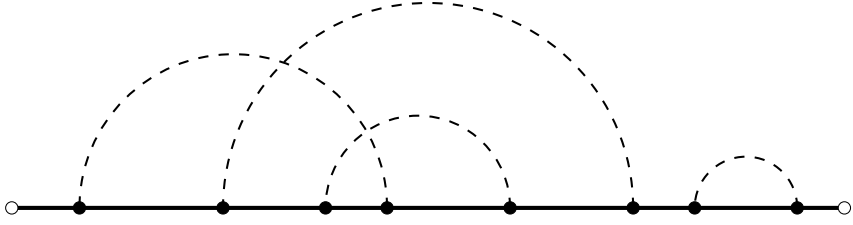}
\caption{A $0$-phonon diagram.}
\label{fig:fig1}
\end{figure}
\begin{figure}
\epsfxsize=0.75\textwidth
\epsfbox{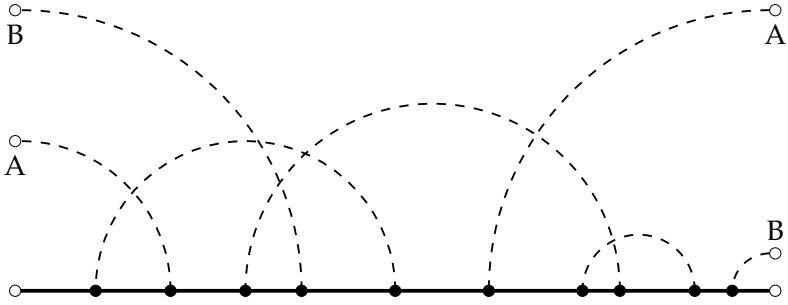}
\caption{A $2$-phonon diagram. Note, that disconnected phonon
lines always appear in pairs (labels A and B), both lines 
in a pair having the same momentum.}
\label{fig:fig2}
\end{figure}
(i) free-electron propagator (solid line)
\begin{equation}
G^{(0)}({\bf p}, \tau_2 - \tau_1 ) \, = \, 
\exp [-\frac{p^2}{2} (\tau_2 - \tau_1)] \; ,
\label{electron}
\end{equation}
(ii) phonon propagator (dashed line)
\begin{equation}
D({\bf q}, \tau_2 - \tau_1 ) \, = \,  
\exp [- \omega_q (\tau_2 - \tau_1)] \; ,
\label{phonon}
\end{equation}
(iii) vertex factor $V({\bf q})$ ascribed to the vertex formed
by a phonon propagator (with the momentum ${\bf q}$) and
two adjacent electron propagators.
External lines of diagrams arise from the operators standing
in Eqs.~(\ref{G}), (\ref{G_N}) and their times and momenta
are defined accordingly. Momenta of the internal lines are free
up to the momentum conservation constraint at each vertex.
To obey this constraint, we choose phonon momenta to be
free, fixing thus the momenta of the electron lines.
Times ascribed to the line ends are subject to the 
chronologization constraint: The time of the left end
is $0$, the time of the right diagram end is $\tau$, times
of the vertexes must increase along the global electron line,
directed from $0$ to $\tau$. Integration over all free parameters of
the diagrams is assumed, within the domains consistent with the 
constraints.

Upon the formulation of the diagrammatic rules, it is reasonable
to introduce the irreducible $N$-phonon Green function, 
$\tilde{G}_N$, which consists only of irreducible diagrams.
[By irreducible diagrams we understand those ones that {\it do not}
contain phonon lines decoupled from the electron.] 
>From this definition it is clear that
the reducible part of the $N$-phonon Green function, 
$G_N - \tilde{G}_N$, is a sum of products of irreducible functions
$\tilde{G}_{N'}$, $N'< N$, and free phonon propagators.
Therefore, the reducible part does not contribute to the 
large-$\tau$ asymptotics of $G_N$ [because of extra factors 
$e^{- \omega_0 \tau}$ coming from the disconnected phonon 
propagators], and, in particular, Eq.~(\ref{asym_N}) holds
true for $\tilde{G}_N$ as well.

It is usefull then to consider the following function:
\begin{equation}
P({\bf k}, \tau) \, = \, G({\bf k}, \tau) + 
\sum_{N=1}^{\infty} \int d{\bf q}_1 \cdots d{\bf q}_N \,
\tilde{G}_N({\bf k}, \tau ; \, {\bf q}_1, \ldots , {\bf q}_N ) 
\label{P}
\end{equation}
[Note, that if $\tilde{G}_N \to G_N$,
this expression would be
singular at $\tau \to 0$ because of the divergence of the integrals for
disconnected phonon propagators]. The function $P$ is readily calculated
by our MC procedure (see next sections) and, according to (\ref{asym}), 
(\ref{asym_N}), satisfies the relation
\begin{equation}
P({\bf k},\tau \gg \omega_0^{-1} ) \; \to \; 
e^{ -E({\bf k}) \tau } \; .
\label{asym_P}
\end{equation}
Here we took into account the completeness relation
\begin{equation}
Z^{({\bf k} )}_0 + \sum_{N=1}^{\infty} \int d{\bf q}_1 \cdots d{\bf q}_N \,
Z^{({\bf k} )}_{N} ({\bf q}_1, \ldots , {\bf q}_N ) \, = \, 1 \; .
\label{complete}
\end{equation}
Eqs.~(\ref{asym_P}), (\ref{complete}) imply that the lowest level
$E({\bf k})$ is non-degenerate. Otherwise one should introduce 
degeneracy factors to the right-hand sides of the equations.

\section{Diagrammatic Quantum Monte Carlo}
\label{sec:3}

In the previous work \cite{PS,PST} it has been shown how to
sum convergent (and arbitrary otherwise) diagrammatic series
numerically without systematic errors. In this section we outline
the basic numerical procedure of evaluating series for various
Green functions described in the previous section [the details
being discussed in the Appendix A], and introduce a number
of estimators, that render the evaluation process significantly
more efficient.

Suppose that we are interested in the function $Q(\{ y\} )$,
which depends on a set of variables $\{ y \}$, and which is 
given in terms of a series of integrals with an ever increasing
number of integration variables:
\begin{equation}
Q(\{ y \}) \, = \, \sum_{m=0}^{\infty} \sum_{\xi_m} \int
dx_1 \cdots dx_m \, {\cal D}_m(\xi_m, \{ y \}, x_1, \ldots , x_m) \; .
\label{main}
\end{equation}
Here $\xi_m$ indexes different terms/diagrams of the same order
$m$. The term $m=0$ is understood as a certain function of $\{ y
\}$.  Both external $\{ y \}$ and internal $\{ x_i \}$ variables
are allowed to be either continuous or discrete; in the latter
case integrals are understood as sums. Diagrammatic MC process
is a numeric procedure based on the Metropolis principle
\cite{Metro} that samples various diagrams in the parameter
space $(\{ y \},m,\xi_m,\{ x \}_m )$ and collects statistics for
$Q(\{ y \})$ in such a way that the final result converges to
the exact answer.  The process has very much in common with the
Monte Carlo simulation of a distribution given by a multi-dimensional 
integral.  Nevertheless, there is an essential difference associated 
with the fact that integration multiplicity in the expansion 
Eq.\ (\ref{main}) is varying.

Summing the series for $Q(\{ y\} )$ is the process of sequential
stochastic generation of diagrams described by functions ${\cal D}_m$,
Eq.~(\ref{main}). The MC process consists of a number of
elementary updates falling into two qualitatively different
classes: (I) those which do not change the type of the diagram
(change the values of arguments in ${\cal D}_m$, but not the function
itself), and (II) those which do change  the
diagram order. The set of elementary updates and their implementation
is problem-specific; the only necessary requirements are
ergodicity, i.e., given two arbitrary diagrams it takes finite
number of updates to transform one to another, and detailed
balance, i.e., diagrams contribute to the statistics according
to ratio of their ${\cal D}$-functions. 
In the Appendix A we describe a set of
updates which we find to work very efficiently. However,
considering enormous freedom in constructing various updates, 
we have little doubts that they may not be further improved.

Though Green function contains complete information associated with 
the polaron spectrum, and the accumulation of the histogram for
$G$ is straightforward, it is reasonable to introduce certain direct
estimators, including one for the Green function itself. 
These estimators substantially enhance the accuracy
of calculations and/or allow collecting more information 
during one MC run (by ``spreading" the data to the different values
of the external parameters). 

\subsection{Estimators for effective mass, group velocity, and
energy}

We start with the family of estimators that
are constructed in accordance with the following standard MC rule. 
Suppose we have some quantity $A$ specified by the diagrammatic expansion
\begin{equation}
A \, = \, \sum_{\nu} \, {\cal D}^{(A)}_{\nu} \; ,
\label{A}
\end{equation}
where ${\cal D}^{(A)}_{\nu}$'s are the diagrams for $A$
which are parametrized by
internal variables denoted by the unified index $\nu$, and 
the summation over $\nu$ is understood as the summation over 
discrete variables and integration over the continuous ones. 
Suppose next that all ${\cal D}^{(A)}_{\nu}$'s are positive
definite and we have a MC process of generation of random
$\nu$ with probability density given by ${\cal D}^{(A)}_{\nu}$.
Then, if some quantity $B$ is specified by a (similar)
diagrammatic expansion
\begin{equation}
B \, = \, \sum_{\nu} \, {\cal D}^{(B)}_{\nu} \; ,
\label{B}
\end{equation}
the estimator for the ratio $B/A$ is given by
\begin{equation}
\frac{B}{A} \, = \, 
\left( \sum_{\mbox{\scriptsize MC}_A \{ \nu \} } Q_{\nu} \right) \, / \,
\sum_{\mbox{\scriptsize MC}_A \{ \nu \}} 1  \; , \; \;
\label{BA}
\end{equation}
\begin{equation}
Q_{\nu} \, = \, \frac{{\cal D}^{(B)}_{\nu} }{{\cal D}^{(A)}_{\nu} } \; .
\label{Q}
\end{equation}
Here $\mbox{\scriptsize MC}_A \{ \nu \} $ means the set of ${\nu}$'s generated
during the MC run.
Commonly, the quantities $A$ and $B$ in Eq.~(\ref{BA}) are, respectively, 
the partition function and an observable. We, however, will use this
relation in a somewhat different context. For one thing, in our case 
$A$ and $B$ can correspond to one and the same Green function, but at 
different values of external parameters (say, momentum or coupling constant).
This way we are able to obtain results for a number of different values 
of the external parameters from a MC process for just one fixed set of
parameters. We can also directly calculate derivatives with respect
to the external parameters. To this end we should analytically take
the corresponding limit from both sides of Eq.~(\ref{BA}). 

Let us obtain estimators for the effective
mass, $m_*$, and group velocity, ${\bf v}({\bf k}) = 
\partial E({\bf k}) / \partial {\bf k}$. 
First we note that ($\tau \to \infty$,~~$\lambda \to 0$)
\begin{equation}
\frac{P({\bf k} + \lambda {\bf \hat{e}}, \tau)}{P({\bf k}, \tau)}
\; \to \; \left\{ 
\begin{array}{ll}
\exp (- \lambda^2 \tau / 2m_*  ) \; , \; & k=0 \, ,  \\
\exp (- \lambda {\bf \hat{e}} {\bf v}({\bf k})  \tau ) \; , \; & k \neq 0 \, ,
\end{array} \right.
\label{PP}
\end{equation}
where ${\bf \hat{e}}$ is a unit vector.
Considering the denominator and the numerator of the left-hand side
of Eq.~(\ref{PP}) as $A$ and $B$ (respectively), we can take advantage
of Eqs.~(\ref{BA}) and (\ref{Q}). The function $Q$ is given by
\begin{equation}
Q \, = \, \prod_j \exp \{- \frac{1}{2} 
[({\bf p}_j +\lambda {\bf \hat{e}} )^2-{\bf p}_j^2](\Delta \tau)_j \} \; .
\label{Q1}
\end{equation}
Here $j$ numerates free-electron propagators of a given diagram 
for $P({\bf k},\tau)$; ${\bf p}_j$ is the momentum corresponding to the 
propagator $j$, and $(\Delta \tau)_j$ is the length of this propagator. 
Eq.~(\ref{Q1}) immediately follows
from the fact that the series for $P({\bf k} + \lambda {\bf \hat{e}}, \tau)$
can be obtained from the series for $P({\bf k}, \tau)$ by
adding the momentum $\lambda {\bf \hat{e}}$ to all free-electron
propagators. As we are interested only in the limit $\lambda \to 0$,
we can expand Eq.~(\ref{Q1}) in powers of $\lambda$:
\begin{equation}
Q \, = \, 1 - \lambda \tau ({\bf \hat{e}} \overline{\bf p}) -
\frac{\lambda^2}{2} \tau + 
\frac{\lambda^2}{2} \tau^2 ({\bf \hat{e}} \overline{\bf p})^2 +
{\cal O}(\lambda^3) \; ,
\label{Q2}
\end{equation}
where $\overline{\bf p}$ is the mean electronic momentum of the
given diagram
\begin{equation}
\overline{\bf p} \, = \, \frac{1}{\tau} 
\sum_j {\bf p}_j \, (\Delta \tau)_j
\; .
\label{P_}
\end{equation}
Comparing Eq.~(\ref{Q2}) with the corresponding expansions of the 
right-hand side of Eq.~(\ref{PP}), we arrive to the following
estimators
\begin{equation}
\left\langle \, \overline{\bf p} \, \right\rangle_{\mbox{\scriptsize MC}}
\; \to \; {\bf v}({\bf k}) \; \; \; \; \; \; \; (\tau \to \infty)
\; ,
\label{v}
\end{equation}
\begin{equation}
1 - \frac{\tau}{3} \, 
\left\langle \, ( \overline{\bf p} )^2 \, 
\right\rangle_{\mbox{\scriptsize MC}}
\; \to \; \frac{1}{m_*} \; \; \; \; \; \; \; (\tau \to \infty)
\; ,
\label{m}
\end{equation}
where $\langle \, \ldots \, \rangle_{\mbox{\scriptsize MC}}$ means
MC averaging in accordance with Eq.~(\ref{BA}).

A special care should be taken for treating the time of the Green
function as an external parameter of the diagrams (in the sense
adopted in this section). The problem is that relations 
(\ref{BA}-\ref{Q}) imply that the internal parameters of diagrams
${\cal D}_\nu ^{(A)}$ and ${\cal D}_\nu ^{(B)}$
have one and the same domain of definition,
otherwise the ratio (\ref{Q}) is not correctly defined. Meanwhile,
the domain of internal times of diagrams directly depends on the external 
time. To circumvent this problem, one can introduce scaled internal times
by simple relation $\tau_i=\tau \tilde{\tau}_i$, where $\tau$ is the
external time (length of diagram in time), $\tau_i$ is an
internal time variable (position in time of an electron-phonon vertex), 
and $\tilde{\tau}_i \in [0,1]$ is the corresponding scaled time 
variable with the domain of definition independent of $\tau$. 

Now it is easy to obtain a direct estimator for the polaron energy.
To this end we start from the relation
\begin{equation}
\frac{P({\bf k}, (1+ \lambda) \tau)}{P({\bf k}, \tau)}
\; \to \; 
e^{- \lambda E({\bf k}) \tau}  \; \; \; \; \; \; (\tau \to \infty) 
\label{PP_2}
\end{equation}
and proceed analogously to Eqs.~(\ref{PP}-\ref{m}). In this case
for the function $Q$ we have
\begin{equation}
Q \, = \, (1+\lambda)^N \left( \prod_j
\exp [- \lambda \frac{p_j^2}{2} (\Delta \tau)_j ] \right)
\left( \prod_s \exp [- \lambda \omega_0 (\Delta \tau)_s ] \right) \; ,
\label{Q3}
\end{equation}
where indexes $j$ and $s$ stand for the electron and phonon propagators,
respectively, and $N$ is the number of integrations over times (or, 
equivalently, number of interaction vertexes) in a given diagram.
Then, in the limit $\lambda \to 0$, we expand the right-hand sides
of Eqs.~(\ref{PP_2}-\ref{Q3}) up to terms proportional to $\lambda$,
and in accordance with Eqs.~(\ref{BA}-\ref{Q}) arrive to the estimator
\begin{equation}
 \frac{1}{\tau} \, 
\left\langle \, 
\sum_j \, \frac{p_j^2}{2} (\Delta \tau)_j \, + \,
\sum_s \, \omega_0 (\Delta \tau)_s \, - N \,
\right\rangle_{\mbox{\scriptsize MC}} \; \to \; \; \;
E({\bf k}) \; \; \; \; \; \; \; \; (\tau \to \infty) \; .
\label{E}
\end{equation}

\subsection{Reweighting}

We will also employ the reweighting technique \cite{FS}, which allows 
one to utilize the statistics being
generated for some given set of external variables, $\xi$, for
calculations at a different set, $\xi'$. In terms of the diagrammatic
Monte Carlo, this technique is based on the relation
\begin{equation}
\sum_{\mbox{\scriptsize MC} \{ \nu \} |_{\xi'}} Q_{\nu} (\xi') \; = \,
\sum_{\mbox{\scriptsize MC} \{ \nu \} |_{\xi}} 
\frac{{\cal D}_{\nu}(\xi')}{{\cal D}_{\nu}(\xi)} \, Q_{\nu} (\xi') \; , 
\label{reweight}
\end{equation}
where $Q_{\nu}$ is any quantity summed over MC statistics. [We omitted
superscript $A$ at ${\cal D}_{\nu}$ since it is not relevant here.] 
The relation (\ref{reweight}) follows from the fact that the 
MC statistics for the set $\xi'$ involves the same (in the sense of
structure and the values of internal parameters) diagrams as the
statistics for the set $\xi$. The difference is only due to a different
probability to generate a diagram with the set $\xi$ rather than $\xi'$.
This difference can be taken into account analytically by the 
corresponding ratio, which immediately leads to (\ref{reweight}).

In our case, typical external parameters are the interaction constant and
the polaron momentum: $\xi = (\alpha, {\bf k})$. The corresponding
ratio of the diagrams is
\begin{equation}
\frac{{\cal D}_{\nu}(\alpha', {\bf k}')}{{\cal D}_{\nu}(\alpha, {\bf k})} 
\, = \, \left( \frac{\alpha'}{\alpha} \right)^{N/2} \prod_j 
\exp \{ -[({\bf k}'-{\bf k})^2 + 2{\bf p}_j ({\bf k}'-{\bf k})]
(\Delta \tau)_j/2 \} \, = \, \left( \frac{\alpha'}{\alpha} \right)^{N/2} 
e^{-[({\bf k}'-{\bf k})^2/2 + 
\overline{\bf p}({\bf k}'-{\bf k})] \tau} \; . 
\label{ratio}
\end{equation}
This relation allows to get many points at different ${\alpha}'$
and ${\bf k}'$, at no extra cost in CPU time, while
performing MC at a given set of ${\alpha}$ and ${\bf k}$
(cf. Ref.~\onlinecite{FS}).

\subsection{Exact estimator for Green's function}

Calculation of the Green function by means of histogram, though is 
simple and natural, involves an apparent shortcoming, associated
with the finite width of the histogram cell. There is always a
competition between the decreasing systematic error by making
the size of the cell smaller, and the increasing statistical accuracy,
which requires increasing the size of the histogram cell. This problem
can be solved by introducing an {\it exact} (free of systematic errors)
estimator for the Green function, as follows from a generic consideration
presented below.

Given some function $A(\xi_0)$ of an external variable/set of variables 
$\xi_0$, specified with the (positive definite) diagrammatic expansion
\begin{equation}
A(\xi_0) \, = \, \sum_{\nu} \, {\cal D}_{\nu} (\xi_0) \equiv 
\int d\xi \sum_{\nu} \, {\cal D}_{\nu} (\xi) \delta (\xi -\xi_0) \; ,
\label{AA}
\end{equation}
[considering a general case, we do not assume that the domain of 
definition of $\nu$ is independent of $\xi$]
and having arranged a MC process of generating configurations $\{\nu , \xi \}$ 
with the probability density proportional to ${\cal D}_{\nu} (\xi)$,
we would like to construct an estimator $a_{\xi_0}$ the average
of which over the MC process gives (up to a global normalization
factor) the function $A(\xi_0 )$. 
Let us look for $a_{\xi_0}$ in the following form
\begin{equation}
a_{\xi_0}(\nu,\xi) \, = \, \left\{ 
\begin{array}{ll}
q(\nu) {\cal D}_{\nu} (\xi_0) / {\cal D}_{\nu} (\xi)  , \; &  \; 
\mbox{if}~\xi \in \Gamma_0~\mbox{and}~{\cal D}_{\nu} (\xi)  \neq 0 \; ,\\
0 \; , \; & \mbox{otherwise} \; .
\end{array} \right.
\label{a_xi_0}
\end{equation}
Here $\Gamma_0$ is some finite domain in the space of variable $\xi$ 
including the point $\xi_0$, $q(\nu)$ is some function to be defined
later. [We adopt a convenient and consistent with the MC procedure
convention that ${\cal D}_{\nu} (\xi)  \equiv 0$, if $\xi$ is out of 
the range of definition of the corresponding diagram.]
>From Eq.~(\ref{a_xi_0}) we have
\begin{equation}
\left\langle \, a_{\xi_0}  \, \right\rangle_{\mbox{\scriptsize MC}} 
\, \equiv  \, 
C \sum_{\nu} \int d \xi \, 
a_{\xi_0}(\nu,\xi) \, {\cal D}_{\nu} (\xi) \, = \,
C \sum_{\nu} \, q(\nu) \, {\cal D}_{\nu} (\xi_0)
\int_{\xi \in \Gamma_0,~{\cal D}_{\nu} (\xi)  \neq 0} d \xi 
\label{a_MC}
\end{equation}
where 
\begin{equation}
C^{-1 } \, = \, 
\sum_{\nu} \int d \xi \, {\cal D}_{\nu} (\xi)
\label{C}
\end{equation}
is the normalization factor for the distribution of the random
pairs $(\nu, \xi)$ induced by the series (\ref{AA}).
>From Eq.~(\ref{a_MC}) it is seen that if we choose
\begin{equation}
q^{-1 }(\nu) \, = \, 
\int_{\xi \in \Gamma_0,~{\cal D}_{\nu} (\xi)  \neq 0} d \xi \; 
\label{q}
\end{equation}
[note that, according to (\ref{a_xi_0}), the definition of $q(\nu)$ 
is relevant only when ${\cal D}_{\nu} (\xi_0)  \neq 0$, 
and that $q^{-1 }(\nu) \neq 0$,
since at least small neighborhood of the point $\xi_0$ contributes
to the integral], then
\begin{equation}
\left\langle \, a_{\xi_0}  \, \right\rangle_{\mbox{\scriptsize MC}} 
\, =  \, C A(\xi_0) \; .
\label{rel_a}
\end{equation}

The particular form of the estimator for the Green function is readily 
obtained by identifying $\xi$ with $\tau$, and noting that the ratio
of diagrams standing in Eq.~(\ref{a_xi_0}) is given in this case by
Eq.~(\ref{Q3}). As the domain of definition of any diagram with
respect to $\tau$ is independent of the diagram 
structure: $\tau \in [0, \infty]$, the factor $q$ in 
Eq.~(\ref{a_xi_0}) is simply proportional to the inverse size of the 
interval $\Gamma_0$. The choice of $\Gamma_0$ for each particular
$\tau_0$ is arbitrary, being a matter of taste and convenience.

\subsection{Improved estimators for phonon statistics}

Collecting statistics for the phonon cloud can be significantly
improved by a trick described below. 

\begin{figure}
\epsfxsize=0.75\textwidth
\epsfbox{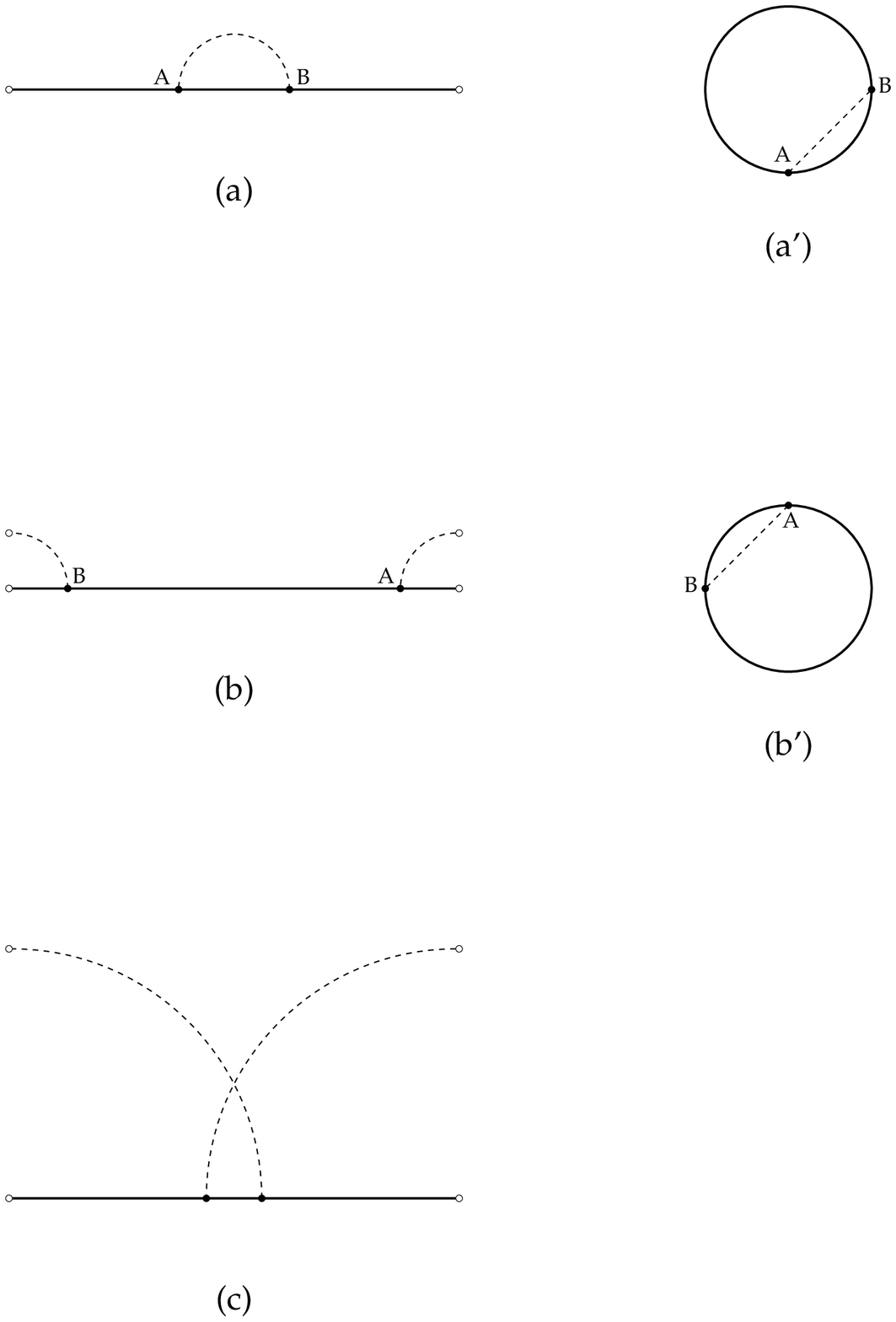}
\caption{Correspondence between the standard and circular representations
of diagrams.}
\label{fig:fig3}
\end{figure}

We start with noting that in the case of $\tau \to \infty$,
which is relevant to the ground-state properties, the set
of all $N$-phonon diagrams possesses a certain symmetry.
To reveal this symmetry, we transform diagrams to
the circular representation by the rules illustrated in 
Fig.~\ref{fig:fig3}.
The transformation involves ``gluing"  the outer ends of the
electron line, as well as the outer ends of pairs of 
(corresponding to each other) external phonon lines. It is easy to
check that the procedure is consistent with the definitions
of propagators, Eqs.~(\ref{electron}, \ref{phonon}).
In the limit of 
large $\tau$, which we are interested in, the probability to
find a phonon propagator with length $> \tau /2$ is
vanishingly small. That is why we may omit orientation,
understanding the time length of phonon propagator as the 
length of the smallest arc between its ends. With the same
accuracy, in the limit $\tau \to \infty$, we may ignore the pairs 
of phonon lines like the one shown in Fig.~\ref{fig:fig3}(c), which do
not have unambiguous circular counterparts. 

It is worth noting that circular diagrams naturally occur
in a finite-temperature technique (cf., e.g., Ref.~\onlinecite{Grabert}), 
where the circumference $\tau$ has a meaning of inverse temperature.

Within the circular representation the symmetry necessary for 
constructing improved estimators is clearly seen. Indeed,
a circular diagram represents a {\it whole class} of
plain diagrams, due to its independence of the position 
of the point corresponding to the ends of the plain diagram.

Thus, having generated a certain  plain diagram and associating it
with the corresponding circular diagram, one effectively produces
a whole class of diagrams to be included into the statistics.

In practice, the procedure is as follows.
Let index $i=1,\dots
, N_e$ label electron propagators in the circular diagram; we thus
split the diagram into $N_e$ pieces each having duration $\Delta
\tau_i $ in time, $\sum_{i=1}^{N_{e}}\Delta_i =\tau $. When the
circular diagram is cut anywhere on the interval $\Delta \tau_i$ we
obtain a contribution to $\tilde{G}_{N_i}({\bf k}, \tau ;
{\bf q}_1^{(i)}, \dots , {\bf q}_{N_i}^{(i)} )$, where $N_i$ is
the number of phonon propagators which are cut along with the electron
propagator on this interval, and $\{ {\bf q}_i \}$ are their momenta.
An estimator for the integrated N-phonon Z-factor is found then
to be
\begin{equation}
Z^{({\bf k})}_{N} \equiv \int \dots \int  \prod_{j=1}^N d{\bf q}_j
Z^{({\bf k} )}_{N}({\bf q}_1, \dots , {\bf q}_{N}) =
\left\langle \sum_{i=1}^{N_e}{\Delta \tau_i \over \tau } \delta_{N_i,N} 
\right\rangle_{MC} \; ,
\label{Z-est}
\end{equation}
i.e., due to time invariance of the circular representation each
interval contributes to the statistics according to its duration
in time. The one-phonon distribution function within the
N-phonon states manifold is given by
\begin{eqnarray}
F^{({\bf k} )}_{N}({\bf q}) \equiv \int \dots \int  \prod_{j=2}^N d{\bf q}_j
Z^{({\bf k} )}_{N}({\bf q},{\bf q}_2, \dots , {\bf q}_{N}) 
 &\equiv & {1 \over N} \sum_{l=1}^N 
\int \dots \int  \prod_{j\ne l}^N d{\bf q}_j
Z^{({\bf k} )}_{N}({\bf q}_1, \dots ,{\bf q }_l ={\bf q}, \dots , {\bf q}_{N}) 
\nonumber \\
 & = & \left\langle \sum_{i=1}^{N_e}{\Delta \tau_i \over \tau } \:
\delta_{N_i,N} \: {1 \over N_i } 
\sum_{j=1}^{N_i} \delta ({\bf q}_j-{\bf q})
\right\rangle_{MC}  \; ,
\label{FN-est}
\end{eqnarray}
Obviously, $Z^{({\bf k} )}_{N} = \int d{\bf q} F^{({\bf k} )}_{N}({\bf q})$.
Summing over all $N$ we obtain the one-phonon distribution function
\begin{equation}
F^{({\bf k} )}({\bf q}) = \sum_{N=1}^{\infty } 
F^{({\bf k} )}_{N}({\bf q})  \; .
\label{F-est}
\end{equation}

\section{Numeric results}
\label{sec:4}

\subsection{Groundstate energy and effective mass}
\begin{figure}
\epsfxsize=0.5\textwidth
\epsfbox{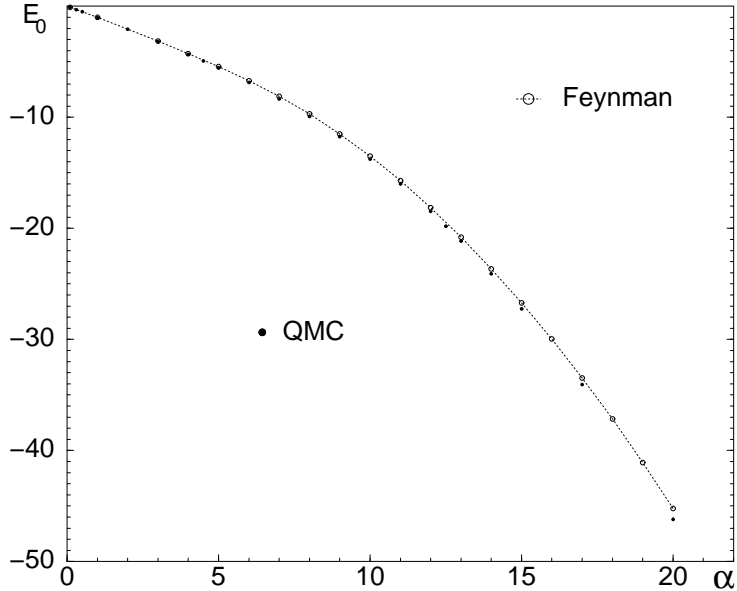}
\caption{Bottom of the polaron band $E_0$ as a function of $\alpha$.
The error bars are much smaller than the point size.}
\label{fig:e_a}
\end{figure}

In Fig.~\ref{fig:e_a} we present our results for the bottom of the band
$E_0$ as a function of $\alpha$ in a wide region of coupling
strengths. The data are compared with Feynman's variational treatment
\cite{Feynman_book} to demonstrate the remarkable
accuracy of the Feynman's approach to the polaron energy.
We thus conclude that just a simple extrapolation between the
second-order perturbative result $E_0=-\alpha-1.26(\alpha/10)^2$ 
and Feynman's strong-coupling variational estimate
$E_0=-\alpha^2/3\pi -2.83$ [such an extrapolation is very close
to Feynman's variational treatment in the whole range of
$\alpha$'s] yields quite satisfactory approximation for
$E_0(\alpha)$.

\begin{figure}
\epsfxsize=0.9\textwidth
\epsfbox{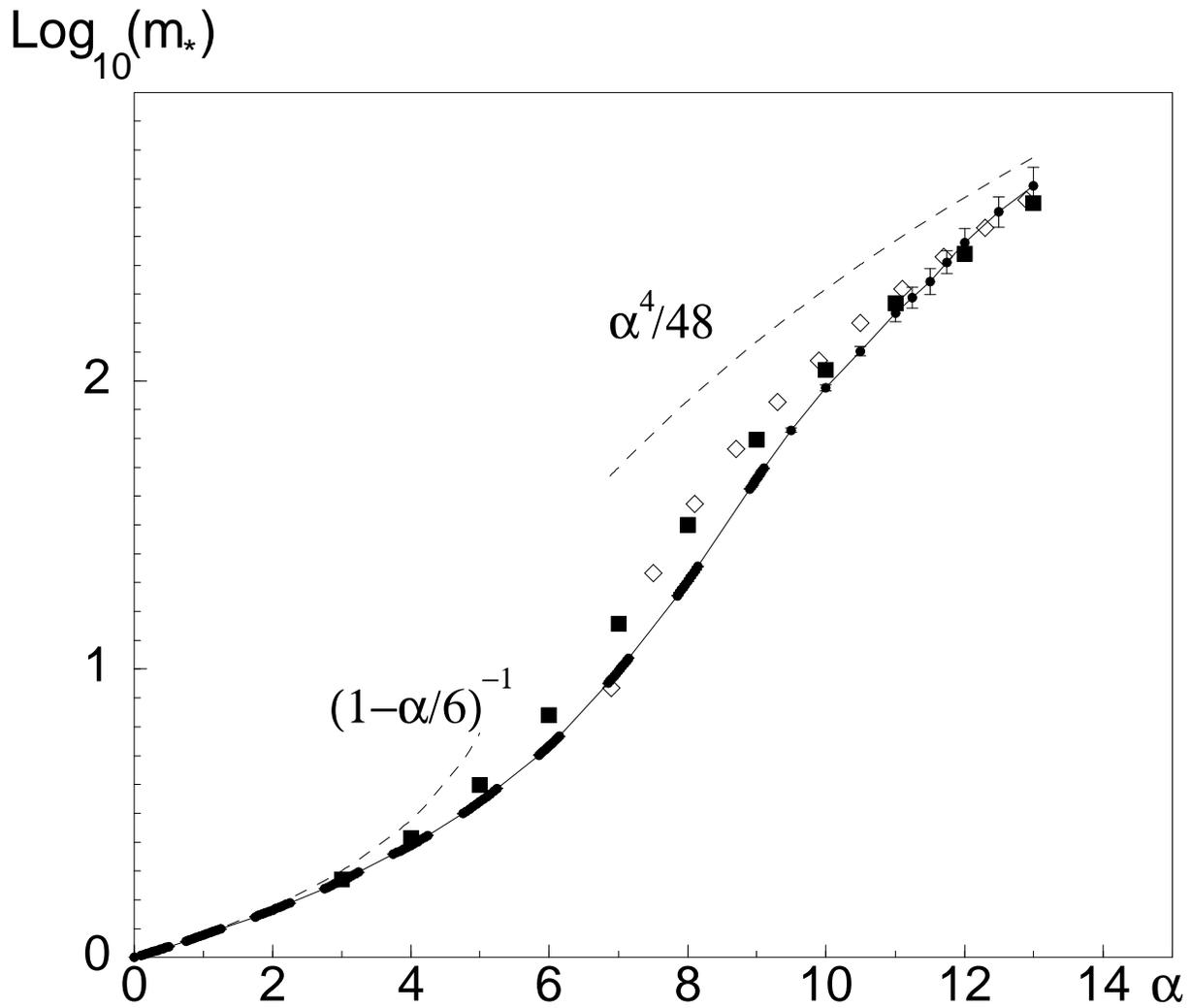}
\caption{Effective mass as a function of coupling parameter. Our MC data
(circles interpolated by solid line) are compared with perturbation theory
and strong-coupling-limit results (dashed lines), Feynman's approach 
(squares), and Feranchuk et. al. 
variational approach (dimonds).}
\label{fig:m_a}
\end{figure}

In Fig.~\ref{fig:m_a} accurate data for the effective mass are presented up
to $m_* \sim 1000$ [For larger $m_*$'s the Fr\"{o}hlich model is
not realistic anyway since the phonon dispersion becomes relevant].
At $\alpha \leq 9$ the statistics were collected at integer values
of $\alpha$ with reweighting (in accordance with the procedure
described in the previous section) to corresponding finite
intervals (see the plot). At $\alpha > 9$ the reweighting
procedure proved to be ineffective.

Let us compare the data for $m_*$ with the small- and strong-coupling analytic
results. At small $\alpha$'s, the formula
$m_*=(1-\alpha/6)^{-1}$, known to coincide with the perturbation
expansion up to the second order \cite{Frohlich,LLP53,Feynman55}, 
works well up to
$\alpha \approx 2$. In contrast to the case of $E_0$, the
strong-coupling limit \cite{LP48} $m_*=\alpha^4/48$ drastically overestimates
the effective mass in the whole range of physically interesting
$\alpha$'s. Feynman's variational technique works better, but still
with a considerable deviation (up to $50$\% )
in the region $5 < \alpha < 10$.

Almost the same degree of accuracy gives variational treatment
by Feranchuk, Fisher, and Komarov \cite{Feranchuk}. An important point
about this treatment, however, is that it suggests a phase
transition from ``light" to ``heavy" polaron at $\alpha$ close
to $7.5$, which should lead to an especially rapid increase of
$m_*$ just after $\alpha=7.5$. In this connection, we note that
our curve $m_*(\alpha)$ is essentially smooth and does not
suggest any sort of phase transition or sharp cross-over.  [The
results of more deep numeric study of the possibility of the phase
transition are presented in the next (sub)sections.]

\subsection{Structure of polaronic cloud}
\label{sec:phonons}

In this subsection we present our results for the electron
Z-factor, or $Z^{({\bf k})}_{0}$, for ${\bf k}=0$ as a function of
the coupling strength in the region of small and intermediate
$\alpha $ (for $\alpha > 10$, the bare electron weight in the
polaron ground state becomes vanishingly small), and for $\alpha
=1$ as a function of momentum up to the end-point. We study
also, how the distribution of phonons in the ground state, $Z_N^{(0)}$,
and the average number of phonons, $\bar{N} =
\sum_{N=1}^{\infty} N Z_N^{(0)}$, evolve with $\alpha$.
Finally, we show how the physics of the end-point is seen in the
transformation of the one-particle distribution function $F_{\bf
k}({\bf q})$, Eq.~(\ref{F-est}), and how it allows to identify
the relevant self-energy diagram.

\begin{figure}
\epsfxsize=0.5\textwidth
\epsfbox{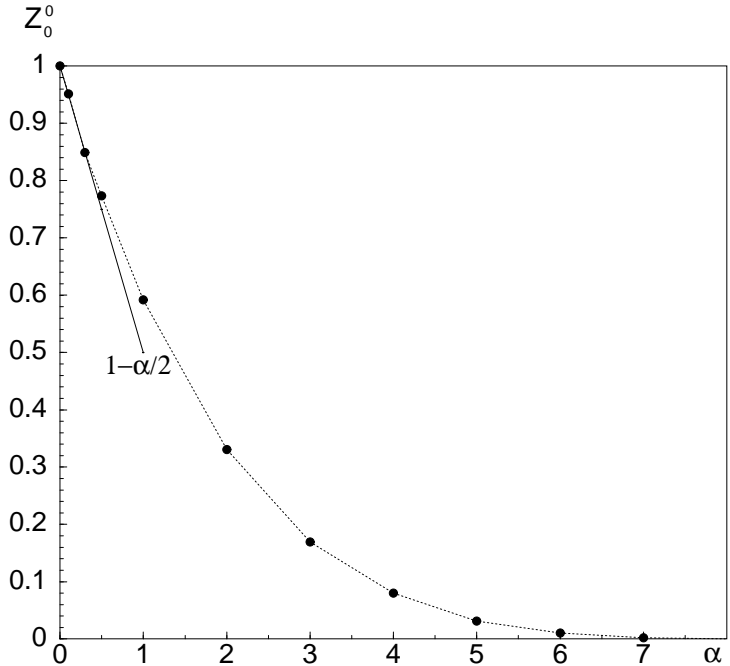}
\caption{The bare-electron $Z^{(0)}_0$-factor for the ground state as a
function of  the coupling strength; filled circles are the MC
data (with error bars smaller than point size), and the solid
line is the perturbation theory result 
(\ref{F1}).}
\label{fig:z_a}
\end{figure}

For small $\alpha $ and ${\bf k} =0$ the leading behavior is
readily obtained from the perturbation theory
\begin{eqnarray}
F^{(0)}_{1}({\bf q})& =& {\sqrt{2} \alpha \over 4 \pi^2 }\: 
{\sin \theta \over [q^2/2+1]^2 } dq \: d\theta \: d\varphi
\nonumber \\
Z^{(0)}_{0} &=& \alpha/2 \nonumber \\
Z^{(0)}_{0} &=& 1-\alpha/2 \;,
\label{F1}
\end{eqnarray}
We have verified that for $\alpha < 1$ the perturbative results
(\ref{F1}) are describing the data rather accurately [see also
Figs.~\ref{fig:z_a}, \ref{fig:n_a}, and dashed curves (connecting
filled circles) in Fig.~\ref{fig:qma1}]. 

The data in Fig.~\ref{fig:z_a} make it clear that perturbation
theory may not be trusted for $\alpha >1$ when the
bare-electron state in the polaron wave function is no longer
the dominant contribution, e.g., $Z_0^{(0)} (\alpha =3) < 0.2$.
The bare electron Z-factor vanishes rather 
rapidly for $\alpha >3$ (the dependence $Z_0^{0}(\alpha )$ 
is faster than exponential) and becomes $<10^{-5}$ for 
$\alpha \ge 10$. We do not attempt to fit the data to the particular
functional dependence since we believe that in the interval $3 <\alpha <10$
the polaron state undergoes a smooth transformation between weak 
and strong coupling limits.

\begin{figure}
\epsfxsize=0.6\textwidth
\epsfbox{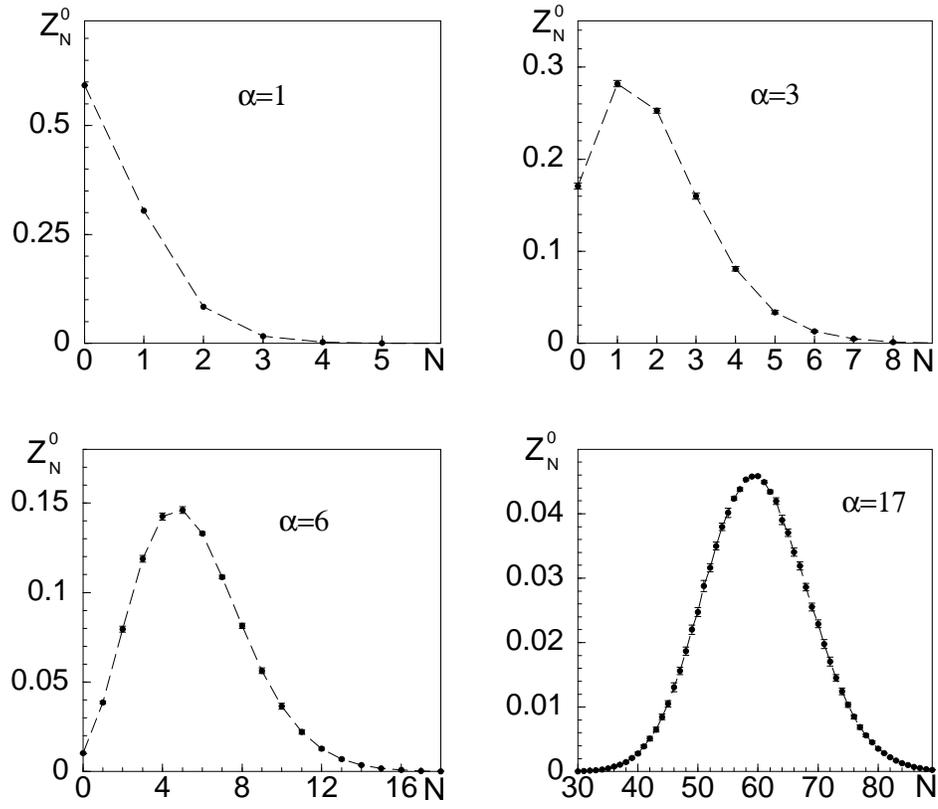}
\caption{Partial contributions of $N$-phonon states
to the polaron ground state for various values of $\alpha$.
Error bars are shown, but are typically smaller than the point size.
(The dashed lines are to guide an eye.)}
\label{fig:z_n}
\end{figure}

In Fig.~\ref{fig:z_n} we show the distribution of multi-phonon states
in the polaron cloud at $k=0$. We see how it gradually  
evolves from the perturbation theory case into the strong-coupling regime.
For $\alpha > 10$ the data may be fit to the Gaussian distribution, but at
smaller values of $\alpha$ the distribution is essentially 
asymmetric - it decays faster for $N> \bar{N}$ than for $N< \bar{N}$. 
We note, that even for $alpha =17$ the phonon number distribution 
is rather large which means that the polaronic cloud is essentially a 
superposition of states with different $N$. The effects resulting from
this fact are outside the scope of the variational $\Psi ^4$-theory
and may, for example, account for the considerable deviation 
of the effective mass discussed above.

It is well known that polaron models often support the self-trapping 
phenomenon, when the ground state changes in a relatively 
narrow interval of parameters from light- to 
heavy-mass state with a sharp increase in the number of phonons contributing
to the polaronic cloud. The same phenomenon was advocated for 
the Fr\"{o}hlich model by a number of authors 
\cite{Gross,Matz,Luttinger,Manka,Manka80,Lepine,Feranchuk}, 
including the statement that more than one stable polaron state exists in 
the region of intermediate $\alpha$.
Clearly, if there were a sharp transformation of the polaronic
ground state, it would have been immediately seen in the phonon statistics.
Such a transformation might be not visible on the energy plot if the
hybridization matrix element between the two competing states is not
small, and  their energy derivatives, $dE/d\alpha$, 
are close to each other at the point of crossing.
However, if we are to speak about {\it different} polaronic states, then
(almost by definition) their structure has to undergo an abrupt
change with $\alpha$. Our data on the bare electron Z-factor $Z_0^0$ and
phonon distribution functions $Z_N^0$ are evolving smoothly with
$\alpha$ and thus prove continuous formation of the self-trapped state.

\begin{figure}
\epsfxsize=0.4\textwidth
\epsfbox{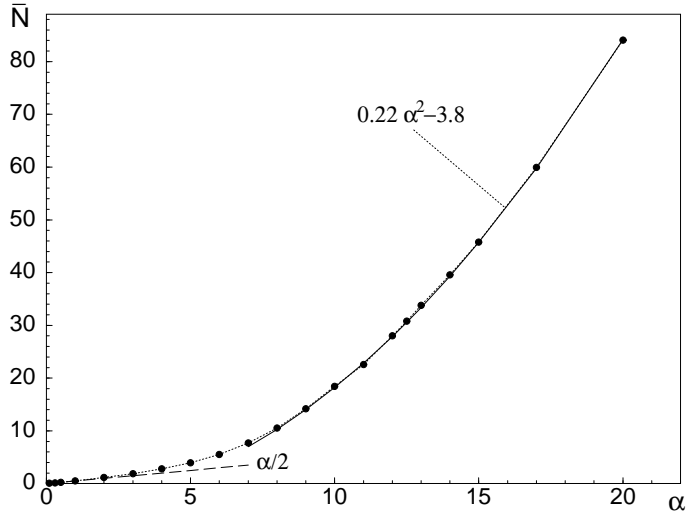}
\caption{The average number of phonons in the polaron ground
state as a function of $\alpha$. Filled circles are the MC data
(error bars are smaller than the point size), the dashed line
is the perturbation theory result 
(\ref{F1}), and the solid line is the parabolic fit for the 
strong coupling limit.}
\label{fig:n_a}
\end{figure}

To further support this conclusion, we plot in Fig.~\ref{fig:n_a} 
the dependence $\bar{N}$ {\it vs} $\alpha$. The crossover between the
perturbative result $\bar{N} 
\sim \alpha /2$ of Ref.~\onlinecite{LLP53}
and the strong-coupling limit, where $\bar{N} \sim 0.22 \alpha^2$, 
demonstrates no sign of the level crossing picture. As a side remark 
we note that the result of Ref.~\onlinecite{LLP53} which predicted 
that perturbation theory for $\bar{N}$ works  well in the intermediate 
range $1< \alpha \le 6$ is not true. In fact, this law breaks down 
along with the perturbation theory.

\begin{figure}
\epsfxsize=0.4\textwidth
\epsfbox{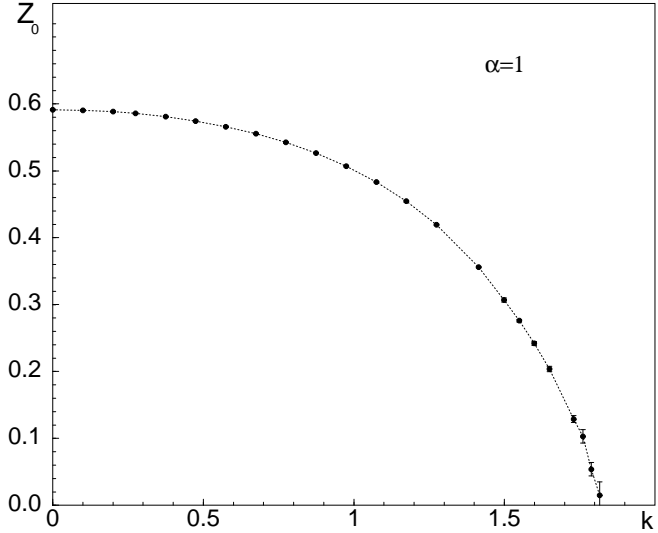}
\caption{The bare-electron $Z$-factor as a function of the 
polaron momentum up to the end point.}
\label{fig:zpa1}
\end{figure}

Consider now the evolution of the polaronic cloud with momentum
as we approach the end point. 
[The dispersion curve $E( k)$ featuring the end point at momentum $k_c$
of the form $E(k \to k_c) = E_0+\omega_0-(k-k_c)^2/2m_c$ was 
calculated in Ref.~\cite{PS}.]
Although the polaronic state is 
stable  for $E(k)-E(0)<\omega_0$ the bare electron weight vanishes
as $k \to k_c$, see Fig.~\ref{fig:zpa1}. From this plot we estimate
$k_c(\alpha =1 ) \approx 1.83$. This figure also makes it clear
to what degree the earlier result that $Z_0^{({\bf k})}$ is 
momentum-independent \cite{Pines} works.

\begin{figure}
\epsfxsize=0.4\textwidth
\epsfbox{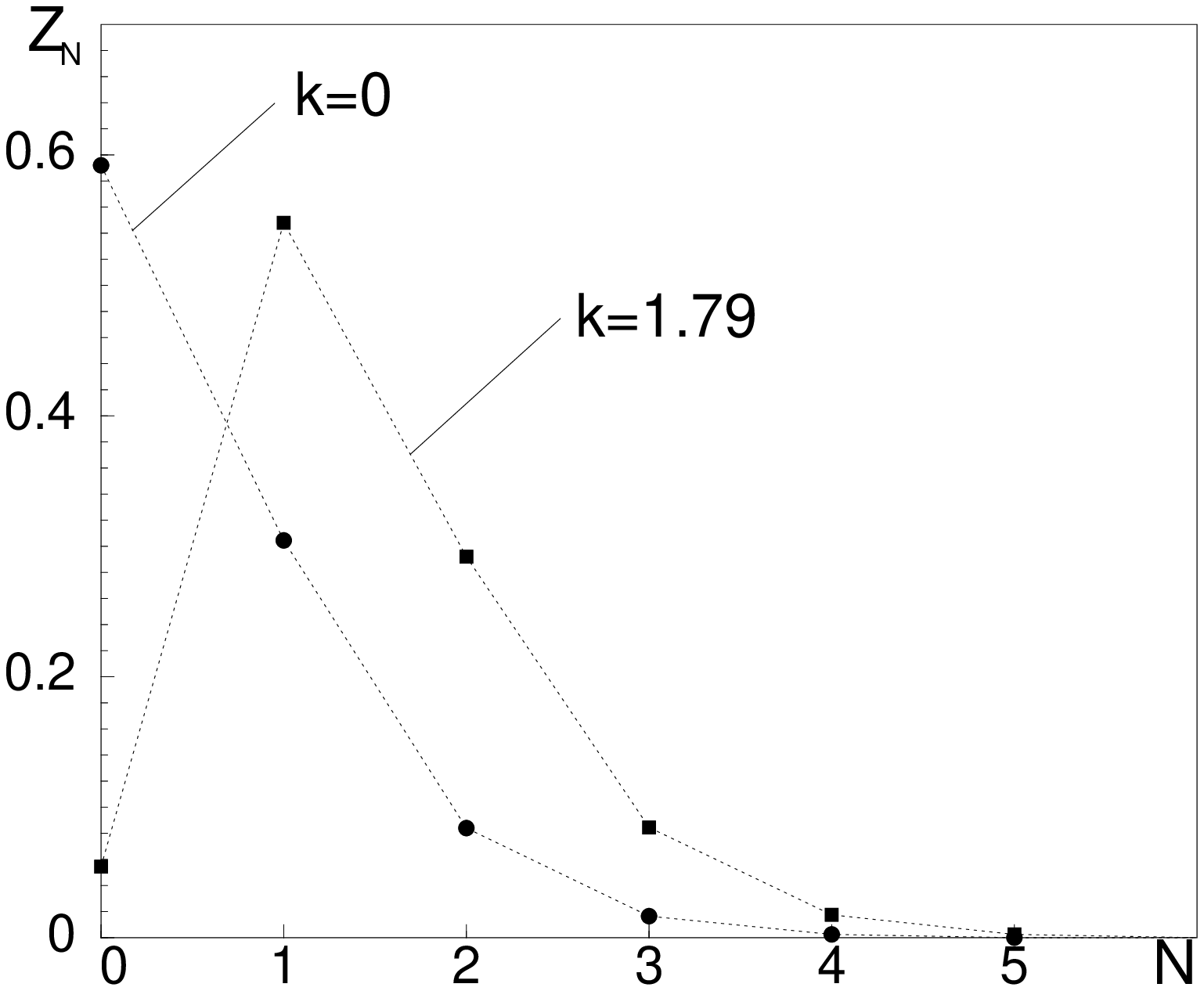}
\caption{Partial weights $Z_N$ of the $N$-phonon states in the structure
of the polaronic state for $\alpha =1$ at the band bottom and near 
the end point. (The dashed lines are to guide an eye.)}
\label{fig:z1e}
\end{figure}

\begin{figure}
\epsfxsize=0.4\textwidth
\epsfbox{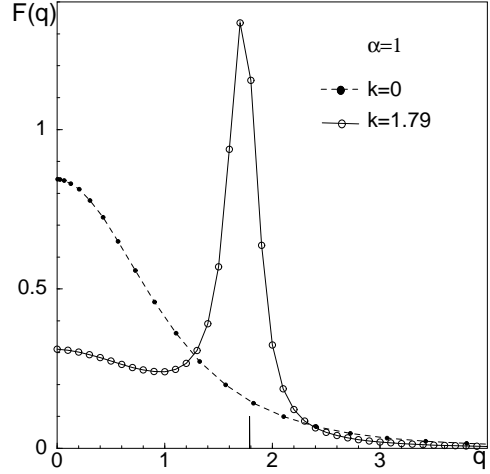}
\caption{Phonon distribution functions in $q$-modulus 
for the ground state (filled circles) 
and close to the end-point for $k=1.79$
(open circles). The momentum $k_c$ is indicated by a bar at the q-axis.
(The lines are just linear interpolations between accurate numeric 
points.) }
\label{fig:qma1}
\end{figure}

With the numerical tools at hand  it is possible to ``visualize" the physics
of the end point. In accordance with the generic Pitaevskii theory of the
end point \cite{Pitaevskii}, in the vicinity of the end point the polaron
can be considered as a (weakly) bound state of phonon, carrying
almost all the momentum of the state, and a polaron with almost zero
momentum. This physics is transparent from the comparison 
between the statistics of $N$-phonon states in the ground state and
at $k \to k_c$ shown in Fig.~\ref{fig:z1e}. Evidently, the two
curves can be matched by shifting the ground-state distribution
by one, i.e.  $Z_N^{(k_c)} \approx Z_{N+1}^{0}$ (the maximum of the 
$Z_N^{(k=1.79)}$ curve is a little depleted because of the 
remaining finite weight $Z_0^0$ since $k<k_c$). It means that
the polaronic state near the end point is a superposition 
of bound states of the phonon with momentum around $k_c$ and 
a polaron at the band bottom. 

\begin{figure}
\epsfxsize=0.4\textwidth
\epsfbox{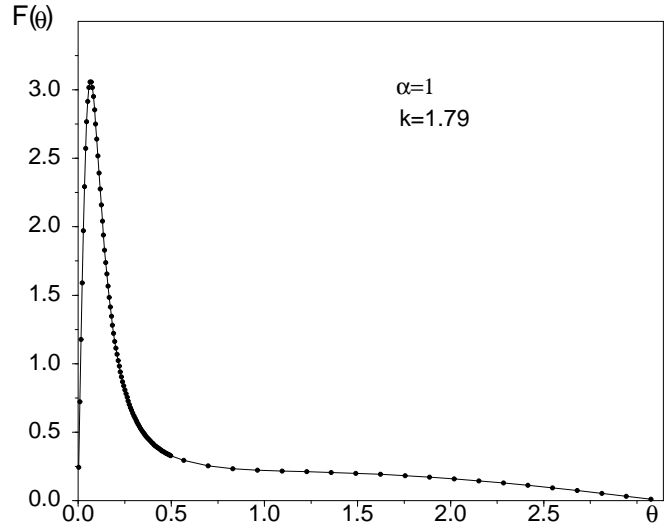}
\caption{Phonon distribution functions in
the angle between the vectors ${\bf q}$ and ${\bf k}$
close to the end-point for $k=1.79$.}
\label{fig:thk18a1}
\end{figure}

Since Fig.~\ref{fig:z1e} does not tell us 
explicitly what are the parameters of the extra phonon present 
in the polaronic cloud at $k \to k_c$, we plot in Figs.~\ref{fig:qma1}
\ref{fig:thk18a1} normalized distribution functions of phonon momenta
(in $\vert q \vert$ and in the angle between $\hat{{\bf q}{\bf k}}$).
It is obvious from these figures that the extra phonon momentum
is concentrated around ${\bf k}$.

 

\section{Spectral analysis}
\label{sec:5}

The spectral function $g_{\bf k} (\omega)$ (\ref{g}) provides
important information about the system since it has poles
(sharp peaks) at frequencies corresponding to stable
(metastable) particle-like states. Besides, since 
the probability of absorption of a free electron 
with the momentum ${\bf k}$ into a polaron state
is proportional to the spectral function,  
the latter can be measured experimentally by 
angle-resolved inverse photoemission
spectroscopy.

In spite of many elaborated
treatments of the properties of the polaron, the knowledge about
high-energy part of the polaron spectrum is mostly limited by
attempts to calculate the spectral density either by perturbation
theory approaches or at strong coupling limit \cite{Devr72}.
As both the Green function asymptotic behavior and the machinery of
estimators provides information about ground state properties
only, the spectral density is indispensable for the study of
excited states of the system.

The spectral analysis, i.e. solving the equation (\ref{Fr}), was
performed by a novel method [detailed description of the method and testing
examples are presented in Appendix II]. 
The most important features of the method are that it avoids distortion 
of equation by nonlinear terms and does not suffer from systematic
errors caused by preassigned discretisation
of the $\omega$-space.

To perform a joint check
of the diagrammatic Monte Carlo approach and the method of
spectral analysis, we compared the spectral densities obtained by our
numeric calculations and by perturbation theory for zero
temperature. The analytic expression for the high-energy part
($\omega>0$) of the spectral density could be  obtained for the
arbitrary interaction potential $V(\mid {\bf q} \mid)$ which
depends on the modulus $\mid {\bf q} \mid$ of the phonon
momentum.  For zero polaron momentum, ${\bf k}=0$, the
the imaginary part of the linear in $\alpha$ self-energy part
$\Sigma(0,\omega>0)$ is
\begin{equation}
\mbox{Im} \Sigma(0,\omega) = 
- \frac{1}{\sqrt{2}\pi}
\sqrt{\omega-1}
\mid V(\sqrt{2(\omega-1)}) \mid^2 \theta(\omega-1).
\label{spa_1} 
\end{equation} 
(Here $\theta$ is the theta-function.)
Then, using the relation 
$g_{\bf 0} (\omega) = -\mbox{Im}G_{{\bf k}=0}(\omega)/\pi$ 
and keeping only linear with respect to $\alpha$ terms one
gets
\begin{equation}
g_{\bf 0} (\omega > 0) = 
\frac{1}{\sqrt{2}\pi^2}
\frac{\sqrt{\omega-1}}{\omega^2} 
\mid V(\sqrt{2(\omega-1)}) \mid^2 \theta(\omega-1).
\label{spa_2} 
\end{equation} 
The expression for the low-energy part ($\omega<0$) of the spectral density 
depends on the specific form of the interaction potential and we consider
the perturbation theory result for the short-range interaction
\begin{equation}
V({\mid \bf q} \mid) \, = \, i \, 
\left( 2 \sqrt{2} \alpha \pi \right)^{1/2} \,  
\frac{1}{\sqrt{q^2+\kappa^2}} \; ,
\label{spa_3}
\end{equation}
which reduces to the Frohlich one when $\kappa \to 0$. The low-energy
part is the delta-functional peak
\begin{equation}
g_{\bf 0} (\omega < 0) = 
\frac{\alpha}{(\kappa + \sqrt{2})^2}
\delta \left( \omega + \alpha \frac{\sqrt{2}}{\kappa + \sqrt{2}} \right). 
\label{spa_4} 
\end{equation} 

\begin{figure}
\epsfxsize=0.85\textwidth
\epsfbox{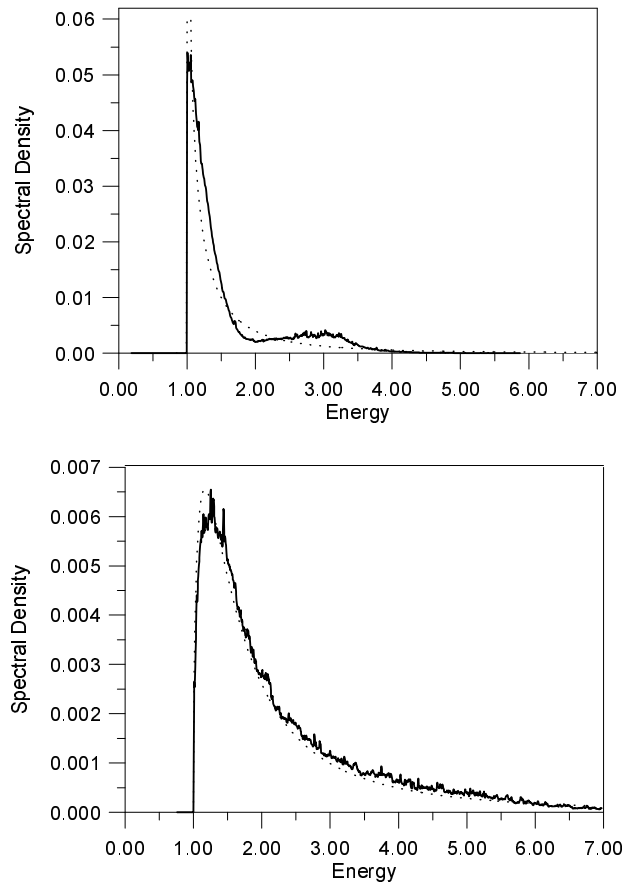}
\caption{The comparison of the numeric results (solid lines) and the 
perturbation theory curves (dashed lines) for the spectral density of 
Frohlich model with $\alpha=0.05$ (upper panel) and the short-range 
interaction model with $\alpha=0.05$ and $\kappa=1$ (lower panel).}
\label{fig:spa_3}
\end{figure}

The comparison of our numeric results for the low-energy part 
of the Frohlich polaron ($\kappa=0$)spectral
density  for $\alpha=0.05$ and Eq.~(\ref{spa_4}) demonstrates
a perfect agreement (with the accuracy $10^{-4}$ for the polaron
energy and Z-factor), whereas our results
for the high-energy part (upper panel in
Fig.~\ref{fig:spa_3}) significantly deviate from 
the analytic curve. 
This is not surprising since for 
Frohlich polaron the perturbation theory expression is diverging 
as $\omega
\to \omega_0$ and, therefore the perturbation theory breaks down.
To test the case when perturbation
theory is obviously valid we set $\kappa=1$ and obtained a perfect
agreement for both the low- and high-energy parts of $g(\omega )$ 
(lower panel in Fig.~\ref{fig:spa_3}). We note that
the high-energy part of $g(\omega )$
is successfully restored by our method despite the fact that
the total weight of the feature is
less than $10^{-2}$ for $\alpha =0.05$.

\begin{figure}
\epsfxsize=0.9\textwidth
\epsfbox{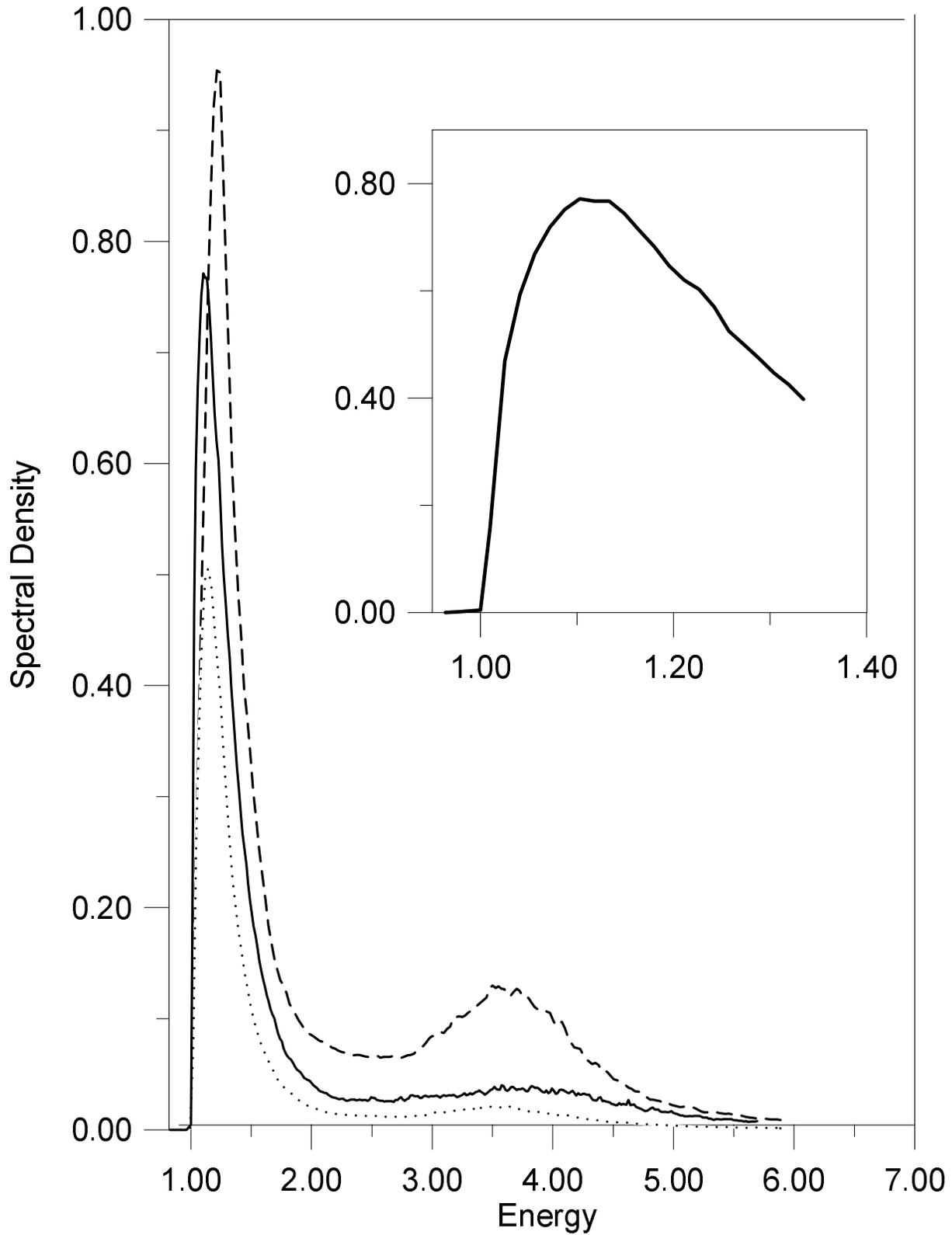}
\caption{The spectral density of Frohlich polaron for $\alpha=0.5$
(dotted line), $\alpha=1$ (solid line) and $\alpha=2$ (dashed line),
with energy counted from the position of the polaron. 
The initial fragment of the spectral density for $\alpha=1$ is shown in the  
insert.}
\label{fig:spa_4}
\end{figure}

One can note that the main deviation of the actual spectrum of
Frohlich polaron from the perturbation theory result is the
extra broad peak in the actual spectral density at $\omega \sim
3.5$. To study this feature we calculated $g(\omega )$
for $\alpha=0.5$, $\alpha=1$, and $\alpha=2$
(see Fig.~\ref{fig:spa_4}). Note, that the peak is seen for
higher values of the interaction constant and its weight grows with
$\alpha$. Near the threshold,
$\omega=1$, the spectral density demonstrates the square-root
dependence $\sim \sqrt{\omega-1}$ (see the insert).

\begin{figure}
\epsfxsize=0.88\textwidth
\epsfbox{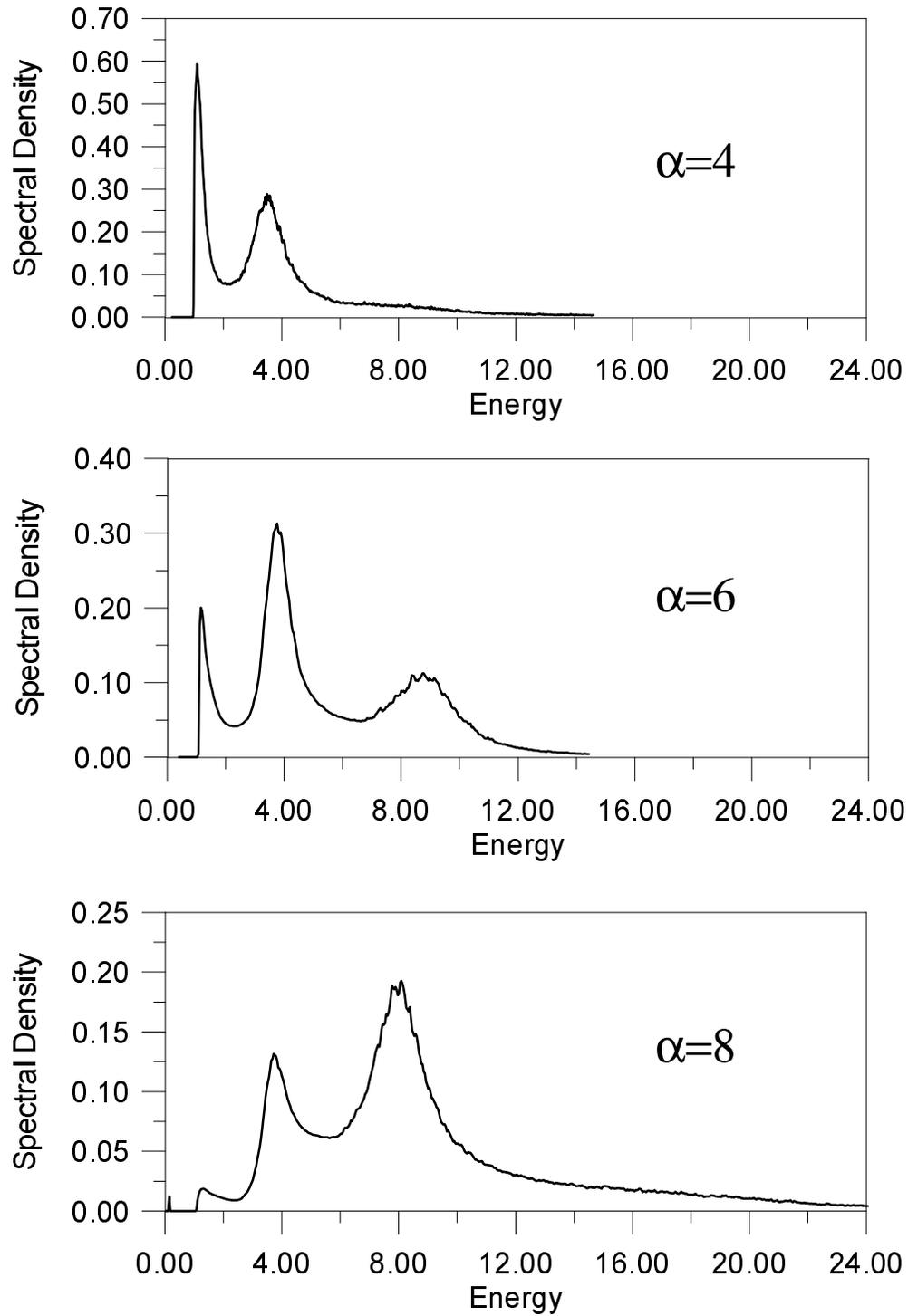}
\caption{Evolution of spectral density with $\alpha $
in the cross-over region from intermediate to strong couplings. (The 
polaron ground state peak is shown only for $\alpha=8$. Note, that the spectral 
analysis still resolves it, despite its very small weight $<10^{-3}$.) 
The energy is counted from the position of the polaron.}
\label{fig:spa_5}
\end{figure}

To trace the evolution of the peak at higher values of
$\alpha$ we calculated the spectral density for $\alpha=4$,
$\alpha=6$, and $\alpha=8$ (see Fig.~\ref{fig:spa_5}). At
$\alpha=4$ the peak at $\omega \sim 3.5$ 
already dominates in the spectral density.
Moreover, a distinct high-energy shoulder appears at $\alpha=4$,
which transforms into a broad peak at $\omega \sim 9$ in the
spectral density for $\alpha=6$. The spectral density for $\alpha=8$
demonstrates further redistribution of the spectral weight between 
different maxima without significant shift of the peak positions.
One can also see that there is a high-energy shoulder
which is, probably, the precursor of another peak which 
would appear for
higher values of the interaction constant.

The excited states of the polaron were studied within the 
frameworks of different 
approaches \cite{GLF62,DHL71,KED69,DSG71,DSG72} 
by calculating optic absorption spectra. The 
light absorption is associated with the transitions from the 
polaron ground state $\phi_0$ with ${\bf k}=0$ and $E=E_0$ 
to the excited states $f$ with $E_f$ which are
characterized by the presence of a finite number of real phonons along
with the polaron. The optic absorption spectrum at the frequency $\omega$
is proportional to the transition probability  
\begin{equation}
P(\omega) = 2 \pi \sum 
\langle \phi_0 \mid \hat{O} \mid f \rangle 
\langle f \mid \hat{O} \mid \phi_0 \rangle 
\delta(E_0-E_f+\omega).
\label{spa_5}
\end{equation}
(Here $\hat{O}={\bf Er}$ is the electric dipole interaction, ${\bf E}$ is the 
electric field.)

It was shown in the weak-coupling limit \cite{GLF62,DHL71}
that the optic absorption spectrum has a broad peak with the onset  
separated from the polaron state by the optic phonon energy. 
Our calculations confirm (see Fig.~\ref{fig:spa_4}) that 
there are no metastable excited states of the polaron in 
the weak-coupling regime. 

On the other hand, in the strong-coupling 
limit the existence of the metastable relaxed excited state (RES), i.e.
the state where the lattice readapts to the new electronic configuration 
and the polaron-lattice system is in the local minimum of the total 
energy, was predicted \cite{KED69,DSG71,DSG72}. 
This state manifests itself as a sharp peak in
the absorption spectrum which is located at the frequency equal to 
the energy difference $E_{\rm RES}-E_0$. 
To check the existence of RES one can study the spectral density 
(\ref{g}) since although the matrix elements of transition probability
(\ref{spa_5}) and spectral density (\ref{g}) are different, both 
functions have to demonstrate sharp peaks at the energies of the 
metastable excited states. From Fig.~\ref{fig:spa_5} we 
conclude that there is no metastable excited state because the width
of the peaks is comparable with the excitation energy, i.e. with the 
distance from the polaron ground state. Moreover, according to
the strong-coupling approaches \cite{KED69}, 
the excitation energy of the RES state is proportional 
to $\alpha^2$, whereas peak positions in $g(\omega )$ with respect to $E_0$
do not change with $\alpha$.   

The variational treatment developed in Ref.~\onlinecite{Feranchuk} suggests
that in a certain region of $\alpha $ there may exist two different
{\it stable } states of the polaron 
[the corresponding equations for variational 
parameters have two solutions]. Our numeric study can shed light on
this situation.

To start with, let us discuss what one could observe would the two states
really exist. If at some point $\alpha=\alpha_*$ there occurs
a level crossing so that the ground state switches from one state
to another, and the two states differ essentially in the number of phonons
and/or in the effective mass, one would expect at the point $\alpha_*$
a sharp change of these quantities. The change should be almost jump-like
in the case of small hybridization between the two states, and look 
like a smooth cross-over otherwise. Even in the case of sufficiently
strong hybridization, one may distinguish between two qualitatively
different cases: (i) the case when the level separation is less than
$\omega_0$ [and thus both states are stable against decay], and (ii) the 
case when the upper level is in the continuum, and therefore is unstable.

Strictly speaking, in the case (ii) one can invoke the second level
only in some quantitative sense, since there is no qualitative difference
between the case (ii) and a situation with only one polaron state. These
quantitative features  could be associated with 
the non-monotonic behavior of the derivatives (with
respect to $\alpha$) of the effective mass and/or the mean number of 
phonons, and, of course, another peak in the spectral density.

Our study of the spectral densities shows that the case (i) is
not realized because we found that $g(0<\omega <1) = 0$.
Therefore, there is no excited stable state in the energy gap
between the ground state energy and incoherent continuum.  Instead, there are
several many-phonon unstable states at energies 
$E_f-E_0 \sim 1$, $\sim 3.5$, and $\sim 8.5$.
One can speculate that these states reveal 
themselves in variational approaches and can be
mistreated as quasi-stable states of the polaron. 
It should be emphasized, however, that the situation does not resemble that 
of the level-crossing at all, since
we do not observe non-monotonic behavior of the derivatives (with respect
to $\alpha$) of the effective mass and/or the mean number of
phonons. 

\section{Acknowledgments}
We are indebted to N.\ Nagaosa for inspiring discussions
and valuable remarks.
This work was supported by the Russian Foundation for Basic Research
(Project 99-02-17288) and by Priority Areas Grants
and Grant-in-Aid for COE research
from the Ministry of Education, Science, Culture and Sports of Japan.

\appendix

\section{Updating procedures}

\subsection{Updates of class I}

Updating procedures of this class are the simplest. They mimic
standard rules of simulating a given distribution function
${\cal D}_m$. In the present case we are dealing with quite a number of
variables having different physical meaning: external variables
$\{ y \}$ include $\tau $, $N$, $\alpha$, and $k$, and internal
variables describe the topology of the diagram (index $\xi_m$),
times of electron-phonon vertices and momenta of phonon
propagators.  From this list of variables follows a set of updates simulating
multi-dimensional distribution ${\cal D}_m$.

\subsubsection{Vertex shift in time}
\label{update1}
We choose at random any interaction vertex inside the graph (we
exclude the diagram closing points which are updated
separately), and change its time position from $\tau_v$ to
$\tau_v'$ on the interval $(\tau_1,\tau_2)$ between the nearest
left- and right-neighbor vertices, i.e., $\tau_1 < \tau_v,\;
\tau_v' < \tau_2$. Let the incoming and outgoing electron
momenta for the selected vertex are ${\bf p}$ and ${\bf p}+{\bf
q}$.  The  normalized probability density to find the vertex at
time $\tau_v'$ is a simple exponential function
\begin{equation}
W(\tau_v') = {\Delta E \, e^{ -(\tau_v'-\tau_1) \Delta E } \over 
1- e^{ -(\tau_2-\tau_1) \Delta E } } \;,
\label{upd1}
\end{equation}
where $\Delta E = E({\bf p})-E({\bf p}+{\bf
q}) \mp \omega_p$ depending on whether the updated vertex 
is the left or right end of the corresponding phonon propagator, 
which allows a trivial solution of the equation 
\begin{equation}
\int_{\tau_1}^{\tau_v'} W(s) ds = r \;,
\label{upd1r}
\end{equation}
in the form
\begin{equation}
\tau_v' = \tau_1-{ \ln 
\left( 1-r[1-e^{ -(\tau_2-\tau_1)\Delta E } ] \right)
\over \Delta E  } \;.
\label{upd1fin}
\end{equation}
Here and below $r$ is the random number homogeneously
distributed on the unit interval. Since the new variable is
selected according to the exact probability density the
acceptance ratio for this update is unity.

\subsubsection{ Change of transferred momentum angle }
\label{update2}
We choose at random any phonon propagator except those attached
to the diagram ends (propagators attached to the diagram ends
appear in {\it pairs} with equal momenta, thus single propagator
updates do not apply to them) and change its momentum 
${\bf q} \to {\bf q}'$ so that 
$\vert {\bf q} \vert = \vert {\bf q}' \vert $. Let the propagator 
connects vertices at times  $\tau_1$ and $\tau_2$. Evaluating 
the average electron momentum between these vertices 
\begin{equation}
<{\bf p}>_{\tau_1, \tau_2} = 
{ \int_{\tau_1}^{\tau_2} {\bf p} (\tau ) d\tau \over
\tau_2-\tau_1 } \;,
\label{pav}
\end{equation}
and introducing vector ${\bf p}_0 = <{\bf p}> _{\tau_1,
\tau_2}+{\bf q}$,
we may write the probability density to find azimuthal and polar
angles $\varphi , \theta $ between vectors ${\bf q}$ and 
${\bf p}_0$  as 
\begin{equation}
W(\varphi , \theta ) \sim \sin (\theta ) \exp \left\{ -{ \tau_2 -\tau_1
\over m } \, p_0 q \cos \theta \right\} \;, 
\label{upd2}
\end{equation}
This result is a trivial consequence of the quadratic dispersion
law for the bare electron spectrum. Clearly, the 
new azimuthal angle is selected
at random ($\varphi = 2\pi r$), and $\cos \theta $ is selected
according to the simple exponential function in complete analogy
with Eqs.~(\ref{upd1}) and (\ref{upd1fin}) up to trivial change
of notations. The acceptance ratio is thus unity.

\subsubsection{ Change of transferred momentum modulus }
\label{update3}
In this procedure propagators are selected as explained in the
previous subsection, but now we change the modulus of the transferred
momentum while keeping the polar and azimuthal angles between
the vectors ${\bf p}_0$ and ${\bf q}$ fixed. The probability density
now reads
\begin{equation}
W(q ) \sim V(q)q^2 \exp \left\{ -{ \tau_2 -\tau_1
\over 2m } \, [q-p_0 \cos \theta]^2 \right\} \sim
\exp \left\{ -{ \tau_2 -\tau_1
\over 2m } \, [q-p_0 \cos \theta]^2 \right\}  \;, 
\label{upd3}
\end{equation}
where we have used explicitly the property of the Fr$\ddot{\rm o}$hlich
model that $V(q)q^2$ is $q$-independent. By tabulating the inverse 
error-function
we ensure fast numerical solution of the equation 
${\rm errf}(z)=r$, or 
$z={\rm errf}^{-1}(r )$, 
and thus generation of
the new value $q=z \sqrt{2m/(\tau_2 -\tau_1)}+p_0 \cos \theta$ with
acceptance ratio unity.

\subsubsection{ Change of diagram structure }
\label{update4}
We select at random any nearest-neighbor pair of vertices inside
the graph (again,  the diagram closing vertices are excluded)
and exchange the assignment of the phonon propagators between
these vertices. Namely, if the original momentum transfer was
${\bf q}_1$ in vertex 1 and ${\bf q}_2$ in vertex 2, we suggest
to change these momenta to ${\bf q}_2$  and ${\bf q}_1$
correspondingly. The acceptance ratio for this procedure depends
on whether we are dealing with the left ($c=1$) or right
($c=-1$) ends of the phonon propagators  
\begin{equation}
R=e^{-\tau [ E({\bf p}+c_1{\bf q_1}-c_2{\bf q_2} )
-E({\bf p}) - \omega_0 (c_1-c_2) ] } \;,
\label{upd4}
\end{equation}
where $\tau $ is the time difference, and ${\bf p}$ is the electron momentum 
between the selected vertices. Clearly, this procedure
effectively changes the topology of bosonic lines while keeping
fixed their momenta.

\subsubsection{ Change of diagram length in time }
\label{update5}
This procedure is done in two variants (almost identical to the
procedure of shifting the vertex position in time). Consider
the case when no artificial potential except the chemical
potential is used. In the first variant we select the new time
difference $\tau $ between the positions of the right diagram end 
at its left nearest neighbor vertex according to the probability density
\begin{equation}
W(\tau' ) = \Delta E \, e^{ -\tau' \Delta E }  \;,
\label{upd5}
\end{equation}
where $\Delta E=E({\bf p})+N_z\omega_0-\mu $, and
${\bf p}$ is the momentum of the last electron propagator,
$\mu$ is the chemical potential, and $N_z$ is the number of phonon
propagators attached to the diagram right end (obviously, $N_z$ is the 
same for the diagram left end).
In the second variant we select new time differences between 
the positions of all nearest-neighbor pairs of vertices. For
each such a pair the probability density is still given by Eq.~(\ref{upd5}),
where in the most general case $N_z$ must be understood as the
number of phonon propagators which are cut when the diagram
is cut anywhere between the selected pair of vertices.  Notice,
that the second variant requires much longer computation
time; thus if the typical diagram order is very large it must be
applied less frequently. In both variants
the acceptance ratio is unity. 

There
is a bottle-neck  in the time decay of the electron $P$-function,
which does not allow efficient sampling of both long-time and short-time
behavior and causes normalization problems at large $\alpha $, namely,
$P(\tau )$ drops to almost zero vaalue at short times, and then climbs 
back to $P \sim 1$ before it settles to the asymptotic decay (\ref{asym_P}).

There  is however a general prescription of how to eliminate
such difficulties by using the so-called ``guiding function'' 
\cite{Ceperley} or
fictitious potential renormalization. This method was successfully
applied recently \cite{instanton} to the problem of
tunneling transition amplitudes, where one is bound to collect
reliable statistics which varies  by orders and orders
of magnitude between different points in time. The idea is to
modify the statistics of suggested diagrams by introducing
the acceptance ratio
\begin{equation}
R=A_{\mbox{\scriptsize fic}}(\tau_{\mbox{\scriptsize new}})/
A_{\mbox{\scriptsize fic}}
(\tau_{\mbox{\scriptsize old}}) \;,
\label{fict}
\end{equation}
and accordingly multiplying all MC estimators in the time domain by
$1/A_{\mbox{\scriptsize fic}}(\tau )$, where the fictitious potential 
$A_{\mbox{\scriptsize fic}}(\tau )$
is arbitrary. Note, that in Eq.~(\ref{fict}) we are dealing with the 
external variable $\tau$ - the diagram length in time.
In the present case the best choice would be
$A_{\mbox{\scriptsize fic}} \sim 1/P(\tau )$. We achieve this goal by 
self-consistently adjusting $A_{\mbox{\scriptsize fic}}$ to 
$1/P_{MC}(\tau )$ after a certain large number of 
updates during the thermalization stage (here $P_{MC}(\tau)$ is the
statistical result for $P(\tau )$). After thermalization stage we 
start collecting new statistics for  $P(\tau )$ and keep 
$A_{\mbox{\scriptsize fic}}$ fixed.

\subsubsection{ Change of  coupling constant }
\label{update6}
Since the diagram weight depends on the coupling constant as
$\alpha^{N_p}$ where $N_p$ is the number of phonon propagators 
in the diagram, all we need to do is to select new value  of $\alpha $
with this power-law probability density. Normalized probability density
is obtained by restricting allowed values of $\alpha$ to a certain
parameter range. Acceptance ratio is unity.

\subsubsection{ Change of  external momentum }
\label{update7}
Given the average electron momentum of the diagram  
$\bar{\bf p} =  <{\bf p}>_{0, \tau }$, see Eq.~(\ref{pav}), with
external momentum ${\bf k}$, we define  vector ${\bf p}_0 =
{\bf k} - <{\bf p}>_{0, \tau} $ and write the probability density to select
new external momentum ${\bf k}'$ as
\begin{equation}
W({\bf k}') \sim \exp \left\{ -{\tau \over 2m}
({\bf k}'-{\bf p}_0)^2 \right\} \;.
\label{upd7}
\end{equation}
As before the new variable is seeded according to this probability density
utilizing the tabulated error-function 
[see subsection \ref{update3} and  Eq.~(\ref{upd3})] 
and thus is always accepted.

One note is in order here. Although one is allowed to change the coupling 
constant and the external momentum in a single MC process, it seems
more efficient to keep these variables fixed instead of spreading
the statistics over some range in the $({\bf k}, \alpha )$ parameter space.
However, the knowledge of the relative weights according to which a given
diagram  contributes to the statistics of various $\alpha $ and ${\bf k}$
may be utilized in collecting statistics for the finite neighborhood
of the point $({\bf k}_0, \alpha_0)$ used in a given MC simulation.
Obviously, reliable results for points other than $({\bf k}_0, \alpha_0)$
are obtained only provided that for typical diagrams the 
relative weights are of order unity. As explained in the text, this 
knowledge is also used in deriving estimators for the effective mass and 
group  velocity of the polaron.

\subsection{Updates of class II}

These updates are in the heart of the method since they
change the diagram order. The generic
rules for constructing them are as follows \cite{PST}.
Let the update $\cal A$ transform a diagram
${\cal D}_m(\xi_m, y, x_1, \ldots , x_m)$ into ${\cal D}_{m+n}(\xi_{m+n}, y, x_1,
\ldots , x_m, x_{m+1}, \ldots , x_{m+n})$, and, correspondingly,
its counterpart $\cal B$ perform the inverse transformation.
For $n$ new variables we introduce vector notation: ${\vec x} =
\{ x_{m+1}, x_{m+2}, \ldots ,  x_{m+n} \}$.  The update
$\cal A$ involves two steps. First, it {\it proposes}
a change, selecting a new diagram, ${\cal D}_{m+n}$, and a
particular value of $\vec x$, which is seeded 
with a certain normalized distribution function $W(\vec x)$. There are no
requirements strictly fixing the form of $W(\vec x)$, but to
render the algorithm most efficient, it is desirable that
$W(\vec x)$ be chosen as close as possible 
to ${\cal D}_{m+n}({\bf x})$, i.e., to the actual statistical probability 
density of $\vec x$ in the new diagram.  Upon proposing the modification,
the update is accepted, with probability, $P_{\mbox{\scriptsize
acc}}(\vec x)$, or rejected.  The update $\cal B$, removing variable
${\bf x}$, is accepted with probability $P_{\mbox{\scriptsize rem}}(\vec x)$).
For the pair of complementary updates
to be balanced, the following Metropolis-like prescription
should be fulfilled \cite{PST}:
\begin{equation}
P_{\mbox{\scriptsize acc}}(\vec x)  \: = \: \left\{ 
\begin{array}{ll}
R(\vec x) / W(\vec x)  ,
\mbox{~~~~~~~ if $R(\vec x) <  W(\vec x) $} \; , \\
\;\;\;\;\;\;1 \; ,  \mbox{~~~~~~~~~~~~~~~~ otherwise} \;\;\; ,
\end{array} \right. 
\label{P_acc}
\end{equation}
\begin{equation}
P_{\mbox{\scriptsize rem}}(\vec x)  \: = \: \left\{ 
\begin{array}{ll}
W(\vec x) / R(\vec x)  ,
\mbox{~~~~~~~ if $R(\vec x) > W(\vec x) $} \; , \\
\;\;\;\;\;\;1 \; ,  \mbox{~~~~~~~~~~~~~~~~ otherwise} \;\;\; ,
\end{array} \right.
\label{P_rem}
\end{equation}
where
\begin{equation}
R(\vec x) \, = \, \frac{p_{\cal B}}{p_{\cal A}} \, 
\frac{{\cal D}_{m+n}(\xi_{m+n}, y, x_1,  \ldots , x_m, {\vec x})}
{{\cal D}_{m}(\xi_m, y, x_1,  \ldots , x_m)}
\label{R}
\end{equation}
and $p_{\cal A}$ and $p_{\cal B}$ are the probabilities of
selecting updates ${\cal A}$ and ${\cal B}$, which, in
principle, may differ.
To solve the polaron problem and account for any possible
diagram it is sufficient to have two pairs 
of complementary processes of type II which are described in
detail below.

\subsubsection{ Adding/removing phonon propagators to the diagram}
\label{update8}
Consider the algorithm for the process increasing the number of internal 
phonon propagators (i.e., excluding those attached to the diagram 
closing points) by one. This update is done in two variants which
differ in the probability densities according to which the new propagator
parameters are suggested.
First we select
the time position $\tau_1$ for the left-hand 
end of the extra phonon propagator. 
This is done by choosing at random (with equal probabilities) one of the 
free-electron propagators, and by taking for $\tau_1$ any time (with
equal probability density) within this propagator. 
Then we select
the transferred momentum and propagator length in time using the 
distribution function 
\begin{equation}
W({\bf q}, \tau )= {\omega_0 \over 4\pi q_0 } 
e^{ -\tau \omega_0 (1+q/q_0)^{2}} 
\label{tau-q}
\end{equation}
where $q_0^2/2=\omega_0$, i.e., we first seed $|q|$ according to
$W_1(q) = =1/[q_0 (1+q/q_0)^{2} ]$
(and isotropic around the point ${\bf q}=0$), and then 
$\tau $ according to 
$W_2(q \vert \tau ) = \omega_0 (1+q/q_0)^{2} e^{ -\tau \omega_0 (1+q/q_0)^{2}}$.
Since the typical length of the phonon propagator in time depends on
how close is the polaron momentum to the dispersion law end-point,
we also use another variant of seeding new variables ${\bf q}$ and 
$\tau $, namely, we factorize the distribution function into
$W({\bf q}, \tau )= W_1(q'=\vert {\bf q}-{\bf k} \vert ) W_3( \tau )$ 
(i.e., isotropic around the point ${\bf k}$), where 
$W_3 (\tau ) = \Omega e^{-\tau \Omega }$ and $\Omega \ll \omega_0$ down to
$\Omega \sim 0.01 \omega_0$ close to the end-point.

We underline, that
the above choices are motivated by the physics of the problem, 
in particular, if the combination $V^2(q)q^2$ was some power law
function of $q$ (e.g., when the interaction vertex is non-singular
at small momentum or even goes to zero as $q \to 0$) one would better
have to choose $W_1(q \to 0) \propto V^2(q)q^2$ to ensure that nowhere
in the accessible parameter region the acceptance ratio (see below)
is singular.

Now the proposing 
stage is completed, and we are ready to perform accept/reject step,
following the above prescription, Eq.\ (\ref{P_acc}). The corresponding
function $W({\vec x})$ (${\vec x} \equiv \{ \tau_1, \tau_2, {\bf q} \}$)
reads (for the first version)
\begin{equation}
W({\vec x}) \, = \, \frac{\omega_0 \, e^{ -\tau \omega_0 (1+q/q_0)^{2}} }
{4 \pi \tau_0 q_0 }  \; , 
\label{W}
\end{equation}
where $\tau_0$ is the length of the free-electron propagator, where
the point $\tau_1$ is selected. As mentioned already, this
form of $W$ is by no means the unique one. Apart from the factor
$p_{\cal B} / p_{\cal A}$ which will be discussed later, the ratio
(\ref{R}) is now completely defined since
\begin{equation}
\frac{{\cal D}_{m+n}(\xi_{m+n}, y, x_1,  \ldots , x_m, {\vec x})}
{{\cal D}_{m}(\xi_m, y, x_1,  \ldots , x_m)} =
{2 \sqrt{2} \pi  \alpha \over (2\pi )^3 } e^{ 
-(\tau_2-\tau_1) [\omega_0
+E(<{\bf p}>_{\tau_1,\tau_2}-{\bf q})-E(<{\bf p}>_{\tau_1,\tau_2})]
} \; .
\label{F/F}
\end{equation}

The algorithm for the process ${\cal B}$ is 
to select at random (with equal probabilities) some phonon
propagator, and, if it is not attached to the diagram end,
with the probabilities given in Eqs.\ (\ref{P_rem}) and  
(\ref{W}) remove it.

To complete the description of the sub-processes $\cal A$ and $\cal B$,
we should define the ratio $p_{\cal B} / p_{\cal A}$. It is quite 
reasonable to select  creation and annihilation procedures
with equal probabilities. At the first glance it might seem that
this immediately leads to $p_{\cal B} / p_{\cal A} = 1$,
but this is not true. The point is that when we select an electron
propagator for placing the point $\tau_1$, we have $N_e$ equal
chances, where $N_e$ is the number of free-electron propagators
in the diagram being modified [denominator of Eq.\ (\ref{R}) ], 
and when we select a phonon propagator for removing, we have $N_{ph}$ 
equal chances, where $N_{ph}$ is the number of phonon propagators 
in the diagram from which we try to remove the propagator 
[numerator of Eq.\ (\ref{R}) ]. These $N_e$ and $N_{ph}$ are 
straightforwardly related to each other:
\begin{equation}
N_{ph} = (N_e +1)/2    \; .
\label{rel}
\end{equation}
We thus get
\begin{equation}
\frac{p_{\cal B} }{ p_{\cal A}}  = \frac{2N_e}{N_e +1} =
\frac{2N_{ph} - 1}{N_{ph}}    \; .
\label{p_ratio}
\end{equation}
(Note a misprint in Ref.~\onlinecite{PS}, where the r.h.s. 
of Eq.~(10)  gives $p_{\cal A}/ p_{\cal B}$ instead of 
$p_{\cal B}/ p_{\cal A}$).
\subsubsection{ Adding/removing a pair of phonon propagators attached 
to diagram ends}
\label{update9}

This update is done in close analogy with the previous one, except
minor changes to which we proceed now. First, we select
time positions
$\tau_1$ and $\tau_2$ for the left- and right-end propagators according 
to the probability densities $W_l(\tau_1) =\Omega e^{-\Omega \tau_1}$ and 
$W_r(\tau_2) =\Omega e^{-\Omega (\tau -\tau_2)}$. In the first
variant of the update $\Omega =\omega_0 $ and in the second 
$\Omega \ll \omega_0$. Also, the phonon momentum is suggested 
using the same distribution 
$W_1(q'=q)$ or $W_1(q'=\vert {\bf q}-{\bf k} \vert )$. 
We thus have 
\begin{equation}
W({\vec x})  =  \frac{\Omega^2  e^{- \Omega (\tau + \tau_1 - \tau_2)}}
{4 \pi q_0 (1+q'/q_0)^2 }  \; , 
\label{WA}
\end{equation}
and
\begin{eqnarray}
& & \frac{{\cal D}_{m+n}(\xi_{m+n}, y, x_1,  \ldots , x_m, {\vec x})}
{{\cal D}_{m}(\xi_m, y, x_1,  \ldots , x_m)} =
{2 \sqrt{2} \pi  \alpha \over (2\pi )^3 } 
e^{ -\omega_0(\tau+\tau_1-\tau_2)} \times  \nonumber \\
& & \left\{
\begin {array}{ll}
e^{ -[E(<{\bf p}>_{0,\tau_1}-{\bf q})-E(<{\bf p}>_{0,\tau_1})]\tau_1
-[E(<{\bf p}>_{\tau_2,\tau}-{\bf q})-E(<{\bf p}>_{\tau_2,\tau })](\tau -\tau_2)
} & \tau_1 < \tau_2 
\\
e^{ -[E(<{\bf p}>_{0,\tau_2}-{\bf q})-E(<{\bf p}>_{0,\tau_2})]\tau_2
-[E(<{\bf p}>_{\tau_2,\tau_1}-2{\bf q})-E(<{\bf p}>_{\tau_2,\tau_1})]
(\tau_1 -\tau_2)
-[E(<{\bf p}>_{\tau_1,\tau }-{\bf q})-E(<{\bf p}>_{\tau_1,\tau })]
(\tau -\tau_1)
} & \tau_1 > \tau_2 
\end{array}
\right.
\; .
\label{FA/FA}
\end{eqnarray}

The algorithm for the inverse procedure is 
to select at random (with equal probabilities)  a pair of
propagators from the list of pairs attached to the diagram end,
and  with the probabilities given in Eqs.\ (\ref{P_rem}) and  
(\ref{WA}) remove it. Since we select procedures inserting and removing
pairs of propagators with equal probabilities, we have
\begin{equation}
\frac{p_{\cal B} }{ p_{\cal A}}  = \frac{1}{N_z +1}   \; .
\label{p_ratioA}
\end{equation}

\section{Method of spectral analysis}

\subsection{General background and outline of the method}

The problem of restoring positive definite 
spectral density function $\rho(\omega)$
from known imaginary-time Green function $G(\tau)$ is 
the problem of solving linear first-type Fredholm equation 
\begin{equation}
\int_{0}^{\infty} e^{-\tau\omega} \rho(\omega) d \omega = G(\tau)  \; ,
\label{ap_sp1}
\end{equation}
where the domain of definition of the functions $G(\tau)$ and $\rho(\omega)$
is $[0,\infty]$.
The normalization of the Green function $G(0)=1$ implies  
the additional constraint
\begin{equation}
\int_{0}^{\infty} \rho(\omega) d \omega = 1 \; .
\label{ap_sp2}
\end{equation}

In a realistic situation the Green function is known at a discrete set of 
times $\{ \tau_i \}, i=1, ..., N$ with some statistic errors
at each time point. 
As is well known, in this case the problem of solving 
Eq.~(\ref{ap_sp1}) belongs to the class 
of ill-posed problems. 
The characteristic feature of the ill-posed problem is that 
the solution of equation (\ref{ap_sp1}) is not unique even when 
statistic errors are
absent, as there is an infinite number of unknown functions
$\tilde{\rho}(\omega)$ satisfying (\ref{ap_sp1}). 
In the case of finite statistic errors one may face a situation when 
the solution of Eq.~(\ref{ap_sp1}) under the constraint (\ref{ap_sp2}) does 
not exist at all.
Therefore, it is natural to formulate the problem as to find an
approximate solution $\rho_{\mbox{\scriptsize min}}(\omega)$ which reproduces 
$G$ at a finite set of times with smallest deviation 
$D_{\mbox{\scriptsize min}}$.
The definition of the measure of deviation depends 
on the method used, and the value of minimal deviation 
$D_{\mbox{\scriptsize min}}$
is determined by the magnitude of statistic errors. 

There are two fundamental difficulties that are inherent to  the
spectral analysis. The first one is the well-known
saw-tooth instability of the linear Fredholm equation of the first type
 -- an approximate solution $\tilde{\rho}(\omega)$ does not reproduce 
the true solution $\rho(\omega)$ even if $\tilde{\rho}(\omega)$ generates the 
Green function 
\begin{equation}
\tilde{G}(\tau) = 
\int_{0}^{\infty} e^{-\omega\tau} \tilde{\rho}(\omega) d \omega
\label{ap_sp3}
\end{equation}
which reproduces $G(\tau)$ with any preassigned accuracy. This difficulty
is treated usually by the regularization method that smoothes the saw-tooth
noise of approximate solution $\tilde{\rho}(\omega)$. 
The idea of the regularization method is to introduce some nonlinearity 
into Eq.~(\ref{ap_sp1}) that imposes constraints on the derivatives of 
$\tilde{\rho}(\omega)$. There are two main drawbacks of this method. 
First, regularization method is unable to restore the spectral density 
which has sharp features. 
Second, due to a distortion of the initial equation by additional 
regularization terms the approximate solution reproduces the function 
$G(\tau)$ with relatively high deviation $D \gg D_{\mbox{\scriptsize min}}$.
Hence, the information from the most representative region of the 
deviations $D \sim D_{\mbox{\scriptsize min}}$ is lost.

The second difficulty inherent to the problem of solving equation
(\ref{ap_sp1}) is that any representation of $\tilde{\rho}(\omega)$ by a 
preassigned discrete set $\{ \rho(\omega_f)\}, f=1,...,M$ is the source 
of uncontrollable systematic errors. For one thing, if the function 
$\rho(\omega)$ contains a sharp feature with a
significant weight at some $\omega'$, which does not match the discrete
set $\{ \omega_f \}$, this feature cannot be reproduced properly and, 
therefore, the rest of the spectral density can be distorted beyond 
recognition. 
Note, that all iteration methods as well as the  methods based on  
solving the nonlinear system of equations use preassigned discretisation of
the $\omega$-space. 

We present a method of solving equation (\ref{ap_sp1}) that avoids distortion 
of equation by nonlinear terms and, thus, probes the most representative
interval of deviations. Besides, the method does not suffer from systematic
errors as it does not involve preassigned discretisation
of the $\omega$-space. The idea of the method is to generate by a
stochastic procedure a (large enough) set of $M$ positive definite 
statistically independent approximate solutions 
$\{ \tilde{\rho}_j(\omega)\}, j=1,...,M$ with deviation measures 
$D_j \sim D_{\mbox{\scriptsize min}}$. 
And then, taking advantage of the linearity of
Eq.~(\ref{ap_sp1}), choose the final solution as the average
\begin{equation}
\rho({\omega}) \, = \, 
M^{-1} \sum_{j=1}^{M} \tilde{\rho}_j(\omega)  \; .
\label{ap_sp4}
\end{equation}
The reason is that while the particular solution $\tilde{\rho}_j(\omega)$ 
possesses the saw-tooth instability, the stochastic character 
of the procedure of particular solution generation should lead to 
averaging out the saw-tooth noise. Note, that the condition 
$\tilde{\rho}_j(\omega) > 0$ and constraint (\ref{ap_sp2})
substantially enhance the convergence of the averaging (\ref{ap_sp4}). 

The method of generation of a particular solution is based on the 
optimization of the deviation
\begin{equation}
D[\tilde\rho] = 
\int_{0}^{\tau_{\mbox{\scriptsize max}}} 
\left| G(\tau) -  \tilde{G}(\tau) \right| G^{-1}(\tau)
d \tau  \; .
\label{ap_sp5}
\end{equation}
Here $\tau_{\mbox{\scriptsize max}}$ is the 
maximal $\tau$ up to which $G(\tau)$ is known. 
The weight function $G^{-1}(\tau)$ is to efficiently utilize 
information from the whole range 
$[0,\tau_{\mbox{\scriptsize max}}]$, even in the case when 
the function $G(\tau)$ decreases by orders of magnitude with $\tau$.
Note, that we use weight function $G^{-1}(\tau)$ rather than
$\tilde{G}^{-1}(\tau)$ to avoid feedback instabilities in generation 
$\tilde{\rho}(\omega)$. 

Our optimization procedure does not involve preassigned 
fragmentation of the $\omega$-space. The number of parameters 
used for parametrization of the spectral density function 
$\tilde{\rho}(\omega)$ is being varied during optimization process, 
so that any spectral function can, in principle, be reproduced within
any preassigned accuracy. The process of generating a particular 
solution involves a random choice of the initial-configuration 
parameters and subsequent optimization of the deviation 
by changing the parameter values, as well as the number of the
parameters. The maximal number of continuous parameters 
and the number of particular solutions $M$
are limited only by the computer performance.

\subsection{Configuration and method of getting independent solution}

We parametrize $\tilde{\rho}$ as a sum
\begin{equation}
\tilde{\rho}(\omega) = \sum_{t=1}^{K} \chi_{\{ P_t \}}(\omega)
\label{ap_sp6}
\end{equation}
of rectangulars $\{ P_t \} =  \{ h_t,w_t,c_t \}$
\begin{eqnarray}
\chi_{\{ P_t \}}(\omega) = \left\{
\begin{array}{ll}
h_t        &, \; \; \; \;  \omega \in [c_t-w_t/2,c_t+w_t/2] \; , \\
0          &, \; \; \; \;  \mbox{otherwise}\; .
\end{array}
\right. 
\label{ap_sp7}
\end{eqnarray}
determined by height $h_t>0$, width $w_t>0$, and center $c_t>0$. 

A configuration 
\begin{equation}
{\cal C} = \left\{  {\{ P_t \}}, \, t=1, ... , K  \right\}
\label{ap_sp8}
\end{equation}
with the constraint 
\begin{equation}
\sum_{t=1}^{K} h_t w_t = 1
\label{ap_sp9}
\end{equation}
defines, according to Eqs.~(\ref{ap_sp6}, \ref{ap_sp3}), the function 
$\tilde{G}(\tau)$ at any time point 
\begin{equation}
\tilde{G}_{{\cal C}}(\tau) = \left\{
\begin{array}{ll}
1         &, \; \; \;   \tau=0 \; , \\ 
2\tau^{-1}
\sum\limits_{t=1}^{K} h_t e^{-c_t \tau} \sinh ( w_t \tau / 2 )
          &, \; \; \;  \tau \ne 0 \; .
\end{array}
\right. 
\label{ap_sp10}
\end{equation}

To express the deviation (\ref{ap_sp5}) as an analytic function 
of the values of $G$ and $\tilde{G}$ at the 
set of times $\{ \tau_i \}, i=1, ..., N$ [where the function $G(\tau)$ 
is known], we use linear interpolation between closest points. 

Note, that the specific type of the functions (\ref{ap_sp7}) is not 
crucial for the general features of the method although simple form 
of analytic expressions (\ref{ap_sp9}, \ref{ap_sp10}) is of value for fast 
performance. 

The procedure of obtaining a particular solution 
$\tilde{\rho}_j(\omega)$ consists of randomly generating some 
initial configuration ${\cal C}_j^{\mbox{\scriptsize init}}$ followed by
nondeterministic sequence of configuration changes until deviation 
satisfies the condition 
\begin{equation}
D[{\cal C}_j^{\mbox{\scriptsize fin}}] < D_u \sim D_{\mbox{\scriptsize min}}
\label{ap_sp11}
\end{equation}
($D_u$ is the upper limiting deviation.)
for final configuration ${\cal C}_j^{\mbox{\scriptsize fin}}$. 
The nondeterministic character of configuration changes is achieved by 
random selection of various elementary updates. 

\subsection{General features of elementary updates}

By elementary update we mean a random change of the configuration,
which is either accepted or rejected in accordance with certain rules.
There are two classes of elementary updates.
The updates of the class I do not alter the number of rectangulars, $K$, 
changing only the values of the parameters from a randomly  
chosen set $\{ P_t \}$. The updates of the class II either add a new 
rectangular with randomly chosen parameters $\{h_{K+1}, w_{K+1}, c_{K+1}\}$,
or remove stochastically chosen rectangular $t$ from the configuration. 
If a proposed change violates constraint (\ref{ap_sp10}) (e.g., a change 
of $w_t$ or $h_t$, or any update of the class II), then the necessary
change of some other parameter set $\{ P_{t'} \}$ is simultaneously
proposed, to satisfy the requirement of the constraint.

The updats should keep parameters of a new 
configuration within domain of definition of configuration ${\cal C}$. 
Formally, the domains of definition of configuration 
(\ref{ap_sp8}) are $\Xi_{h_t} = [0,\infty]$, $\Xi_{c_t} = [0,\infty]$, 
$\Xi_{w_t} =[0,2c_t]$, and $\Xi_K \in [1,\infty]$. However, for the 
sake of faster convergence, we reduce domains of definition. 

As there is no general {\it a priori} prescription for choosing reduced 
domains of definition, the rule of thumb is to start with 
maximal domains and then, after some rough solution is found, reduce 
the domains to reasonable values suggested by this solution.
In particular, since the probability to propose a change of any 
parameter of configuration is proportional to $K^{-1}$, it is natural 
to restrict maximal number of rectangulars 
$\Xi_K \in [1,K_{\mbox{\scriptsize max}}]$ by 
some large number $K_{\mbox{\scriptsize max}}$. To forbid rectangulars 
with extremely small 
weight, which contribution to $\tilde{G}(\tau)$ is less than statistic 
errors of $G(\tau)$, one can impose the constraint
$h_t w_t \in [S_{\mbox{\scriptsize min}},1]$, with 
$S_{\mbox{\scriptsize min}} \ll K_{\mbox{\scriptsize max}}^{-1}$. 
When there is some preliminary knowledge that overwhelming majority
of integral weight of the spectral function $\rho(\omega)$ is in a range 
$[\omega_{\mbox{\scriptsize min}}, \omega_{\mbox{\scriptsize max}}]$, one can 
restrict the domain of definition of the parameter $c_t$ by 
$\Xi_{c_t} = [\omega_{\mbox{\scriptsize min}}, 
\omega_{\mbox{\scriptsize max}}]$. Then, to reduce the phase space 
one can choose $\Xi_{h_t} = [h_{\mbox{\scriptsize min}}, \infty]$ and 
$\Xi_{w_t} = \left[ w_{\mbox{\scriptsize min}},
\min \left\{ 2(c_t-\omega_{\mbox{\scriptsize min}}), 
2(\omega_{\mbox{\scriptsize max}}-c_t) \right\} \right]$.

While the initial configuration, the update type, and the parameter to 
be altered are chosen stochastically, the variation of the 
values of the parameters relevant to the update is optimized to
maximize the decrease of ${\cal D}$.
Each elementary update of our optimization procedure (even that of the  
class II) is organized as a proposal to change some continuous parameter 
$\xi$ by randomly generated $\delta \xi$ in a way that the new 
value belongs to $\Xi_{\xi}$.
Although proposals with smaller values of $\delta \xi$
are accepted with higher probability it is important, for the sake of better 
convergence, to propose sometimes changes $\delta \xi$ that probe 
the whole domain of definition $\Xi_{\xi}$. To probe all scales of 
$\delta \xi 
\in [\delta \xi_{\mbox{\scriptsize min}} , \delta \xi{\mbox{\scriptsize max}}]$
we generate $\delta \xi$ with the probability density function 
$\cal P \sim (  
\max(| \delta \xi_{\mbox{\scriptsize min}}| , 
     | \delta \xi_{\mbox{\scriptsize max}}|) /  | \delta \xi | 
)^{\gamma}$, 
where $\gamma \gg 1$.

Calculating the deviation measures $D(\xi)$, $D(\xi + \delta \xi)$, 
$D(\xi + \delta \xi/2)$, and 
searching for the minimum of the parabolic interpolation,
we find an optimal
value of the parameter change 
\begin{equation}
\delta \xi_{\mbox{\scriptsize opt}} = - B/2A ,
\label{ap_sp12}
\end{equation}
where 
\begin{equation}
A = 
2 ( D(\xi + \delta \xi) - 2D(\xi + \delta \xi /2) + D(\xi)) (\delta\xi)^{-2},
\label{ap_sp13}
\end{equation}
and
\begin{equation}
B = 
(4D(\xi + \delta \xi/2) - D(\xi + \delta \xi) - 3D(\xi))\delta\xi .
\label{ap_sp14}
\end{equation}
In the case $A>0$ and $\xi_{\mbox{\scriptsize opt}} \in \Xi_{\xi}$
we adopt as the update proposal
$\tilde{\delta\xi}$ one of the values $\delta\xi$, $\delta\xi/2$, or 
$\delta\xi_{\mbox{\scriptsize opt}}$ for which the deviation measure 
$D(\xi+\tilde{\delta\xi})$ is the smallest. Otherwise, if the parabola 
minimum is outside $\Xi_{\xi}$, 
one has to compare only deviations for $\delta\xi$ and $\delta\xi/2$.

\subsection{Global updates}

The updating strategy has to provide efficient minimization of the 
deviation measure until criterion (\ref{ap_sp11}) is satisfied. 
It is highly inefficient to accept only those proposals
that lead to the decrease of deviation, since, in a general case,
there is an enormous
number of deviation local minima 
$D_{\mbox{\scriptsize loc}}[{\cal C}] > D_{u}$. 
As we observed it in practice, 
these multiple minima drastically slow down (or even freeze) the process.

To optimize escape from a local minimum, one has to provide 
a possibility of reaching a new local minimum with lower deviation 
through a sequence of less optimal configurations. It might seem
that the most natural way of doing this would be to 
accept sometimes (with low enough probability) the updates leading
to the increase of the deviation.
However, this simple strategy turns out to be impractical. The 
reason is that the density of configurations per interval of 
deviation sharply increases with $D$. So that the 
acceptance probability for a deviation-increasing update
should be fine-tuned to the value of $D$. Otherwise, the 
optimization process will be either non-convergent, or
ineffective [if the acceptance probability is, correspondingly, 
either too large, or too small in some region of $D$].

A way out of the situation is to perform some sequence of $T$ 
{\it temporary} elementary updates of a configuration ${\cal C}(0)$ 
\begin{equation}
{\cal C}(0) \to {\cal C}(1) \to ... \to 
{\cal C}(r) \to {\cal C}(r+1) \to ... \to {\cal C}(T) \; ,
\label{ap_sp15} 
\end{equation}
where the proposal to update the configuration 
${\cal C}(r) \to {\cal C}(r+1)$ is (temporary) accepted 
with the probability
\begin{eqnarray}
{\cal P}_{r \to r+1} = \left\{
\begin{array}{ll}
1          &, \; \; \; \;  D[{\cal C}(r+1)] < D[{\cal C}(r)] \; ,\\
f \left( D[{\cal C}(r)]/D[{\cal C}(r+1)] \right) &, \; \; \; \; 
D[{\cal C}(r+1)] > D[{\cal C}(r)] \; .
\end{array}
\right.
\label{ap_sp16}
\end{eqnarray}
(Function $f$ satisfies boundary conditions $f(0)=0$ and $f(1)=1$.)
Then we choose out of the configurations $\{ {\cal C}(r) \}$ (\ref{ap_sp15})
the one with minimal deviation and, if it is different from ${\cal C}(0)$,
declare it to be the result of the global update, or, if this 
configuration turns out to be just ${\cal C}(0)$, reject the update.

We choose the function $f$ in the form
\begin{equation}
f(x) = x^{1+d} \; \; \; \; \;  (d>0) \; ,
\label{ap_sp17}
\end{equation}
which leads to comparatively high probabilities to accept small 
increases of deviation measures and hampers significant 
enlargements of deviation. Empirically, we found out that the 
global update procedure is most effective if one keeps 
parameter $d = d_1 \sim 0$ at the first $T_1$ steps of sequence (\ref{ap_sp15}) 
(to leave local minimum) and then changes this parameter to a value
$d = d_2 \gg 1$ for the last $T-T_1$ elementary updates (to decrease 
the deviation measure). In our 
algorithm the values $T \in [1,T_{\mbox{\scriptsize max}}]$, 
$T_1 \in [1,T]$, $d_1 \in [0,1]$, and 
$d_2 \in [1,d_{\mbox{\scriptsize max}}]$ were stochastically chosen for 
each global update run. 

\subsection{Final solution and refinement}

After a set of $M$ configurations 
\begin{equation}
\left\{ C_j^{\mbox{\scriptsize fin}}, j=1,...,M
\right\}
\label{ap_sp18}
\end{equation}
that satisfy the criterion (\ref{ap_sp11}) 
is produced, the solution (\ref{ap_sp4}) 
is obtained by summing up the rectangulars (\ref{ap_sp7},\ref{ap_sp18}).

We, however, employ a more elaborated procedure, which we 
call refinement. Namely, we use the set (\ref{ap_sp18}) as a source of 
$M_{\mbox{\scriptsize ref}}$ new independent starting 
configurations for further optimization. These starting configurations 
are generated as a linear combinations of randomly chosen members 
of the set (\ref{ap_sp18}) with stochastic weight coefficients.
Then, the refined final solution is represented as the 
average (\ref{ap_sp4}) of $M_{\mbox{\scriptsize ref}}$ particular solutions 
resulting from the optimization procedure.

The main advantage of such a trick is that initial configurations 
for optimization procedure now satisfy the criterion (\ref{ap_sp11}) from the 
very beginning and, thus, upper limiting deviation $D_u$ can be 
considerably reduced. Moreover, as any linear combination 
of sufficiently large number $R$ of randomly chosen parent configurations 
$\left\{ C_{\eta}^{\mbox{\scriptsize fin}}, \eta=1,...,R \right\}$
smoothes the saw-tooth noise, 
the deviation of a summary configuration  
$C_{\mbox{\scriptsize ref}}^{\mbox{\scriptsize fin}}$ is normally lower 
than that of each additive one. 

\subsection{Elementary updates of class I}

{\it (A) Shift of rectangular}. Change the center $c_t$ of a 
randomly chosen rectangular $t$. The continuous parameter for optimization
(\ref{ap_sp12}-\ref{ap_sp14}) is $\xi=c_t$ which is 
restricted by domain of definition 
$\Xi_{c_t} =
[\omega_{\mbox{\scriptsize min}}+w_t/2 ,
\omega_{\mbox{\scriptsize max}}-w_t/2]$.

{\it (B) Change of width without change of weight}. Alter the width $w_t$ of a 
randomly chosen rectangular $t$ without change of the rectangular weight
$h_t w_t= const$ and center $c_t$. The continuous parameter for optimization 
is $\xi=w_t$ which is restricted by 
$\Xi_{w_t} = \left[ w_{\mbox{\scriptsize min}},
\min \left\{ 2(c_t-\omega_{\mbox{\scriptsize min}}), 
2(\omega_{\mbox{\scriptsize max}}-c_t) \right\} \right]$.

{\it (C) Change of weight of two rectangulars}. Change the heights of two 
rectangulars $t$ and $t'$ (where $t$ is a randomly chosen and $t'$ is either 
randomly chosen or closest to $t$ rectangular) without change of 
widths of both rectangulars. Continuous parameter for optimization
is the variation of the rectangular $t$ height $\xi=h_t$. To restrict
the weights of chosen rectangulars to $[S_{\mbox{\scriptsize min}},1]$ and
preserve the total normalization (\ref{ap_sp2}) this update suggests 
to change $h_t \to h_t+\delta \xi$ and 
$h_{t'} \to h_{t'} - \delta \xi w_{t'} / w_t$ with $\delta \xi$ 
confined to the interval
\begin{equation}
S_{\mbox{\scriptsize min}}/w_t - h_t < \delta \xi
< (h_{t'} - S_{\mbox{\scriptsize min}}/w_{t'}) w_t / w_{t'}
\label{ap_sp19}
\end{equation}

\subsection{Elementary updates of class II}

{\it (D) Adding  a new rectangular}. To  add a new rectangular one has to 
generate some new set 
$\{ P_{\mbox{\scriptsize new}} \} =   
\{ h_{\mbox{\scriptsize new}} , w_{\mbox{\scriptsize new}}  , 
c_{\mbox{\scriptsize new}} \}$ and reduce 
the weight of some other rectangular $t$ (either randomly chosen or closest)
in order to keep the normalization 
condition (\ref{ap_sp2}). The reduction of the rectangular weight $t$ is obtained 
by decreasing its height $h_t$. 

The center of the new rectangular is selected at 
random according to
\begin{equation}
c_{\mbox{\scriptsize new}} = 
(\omega_{\mbox{\scriptsize min}} + w_{\mbox{\scriptsize min}}/2) +
(\omega_{\mbox{\scriptsize max}} - \omega_{\mbox{\scriptsize min}} 
- w_{\mbox{\scriptsize min}}) r
\label{ap_sp20}
\end{equation}
As soon as the value $c_{\mbox{\scriptsize new}}$ is generated, 
the maximal possible width of a new rectangular is given by
\begin{equation}
w_{\mbox{\scriptsize new}}^{\mbox{\scriptsize max}} = 
2 \min (\omega_{\mbox{\scriptsize max}} - c_{\mbox{\scriptsize new}} , 
        c_{\mbox{\scriptsize new}} - \omega_{\mbox{\scriptsize min}} ).
\label{ap_sp21}
\end{equation}

Continuous parameter for optimization 
$\delta \xi = h_{\mbox{\scriptsize new}} w_{\mbox{\scriptsize new}}$
is generated to keep weights of both new rectangular and rectangular $t$
larger than $S_{\mbox{\scriptsize min}}$
\begin{equation}
\delta \xi =  S_{\mbox{\scriptsize min}}
+ r (h_t w_t -  S_{\mbox{\scriptsize min}})
\label{ap_sp22}
\end{equation}
Then, the value of the new rectangular height $h_{\mbox{\scriptsize new}}$
for given $\delta \xi$ is generated to keep the width of new 
rectangular within the limits 
$[w_{\mbox{\scriptsize min}} , 
w_{\mbox{\scriptsize new}}^{\mbox{\scriptsize max}}]$
\begin{equation}
h_{\mbox{\scriptsize new}} = 
\delta \xi / w_{\mbox{\scriptsize new}}^{\mbox{\scriptsize max}} +
r ( \delta \xi / w_{\mbox{\scriptsize min}}
- \delta \xi / w_{\mbox{\scriptsize new}}^{\mbox{\scriptsize max}}). 
\label{ap_sp23}
\end{equation}

{\it (E) Removing a rectangular}. To remove some randomly chosen rectangular 
$t$, we enlarge the height $h_{t'}$ of some another (either randomly chosen or 
closest) rectangular $t'$ according to condition (\ref{ap_sp2}). Since such 
procedure does not involve continuous parameter for optimization,
we unite removing of rectangular $t$ with the shift procedure {\it (A)}
of the rectangular $t'$. Then, the proposal is the configuration with
the smallest deviation measure. 

{\it (F) Splitting a rectangular}. This update cuts some rectangular $t$ into 
two rectangulars with the same heights $h_t$ and widths 
$w_{\rm new_1}=w_{\mbox{\scriptsize min}} +
 r (w_t-w_{\mbox{\scriptsize min}})$ 
and 
$w_{\rm new_2}= w_t - w_{\rm new_1}$. 
Since removing a rectangular $t$ and adding of two new glued rectangulars 
does not change the spectral function we introduce the continuous
parameter $\delta \xi$ which describes the shift of the  center of a new 
rectangular with the smallest weight. Second rectangular is shifted into
opposite direction to keep the center of gravity of two rectangulars 
unaltered. The domain of definition $\Xi_{\xi}$ obviously follows from 
the parameters of the new rectangulars.

{\it (G) Gluing  rectangulars}. This update glue two (either randomly chosen 
or closest) rectangulars $t$ and $t'$ into single new rectangular with the 
weight 
$h_{\mbox{\scriptsize new}} w_{\mbox{\scriptsize new}} = 
 h_t w_t + w_{t'} h_{t'}$ and width
$w_{\mbox{\scriptsize new}} = ( w_t + w_{t'} ) / 2$. 
The initial center of the  new rectangular $c_{\mbox{\scriptsize new}}$
corresponds to the center of gravity of rectangulars $t$ and $t'$. 
We introduce a
continuous parameter by simultaneously shifting the new rectangular.

\subsection{Tests}

\begin{figure}
\epsfxsize=0.95\textwidth
\epsfbox{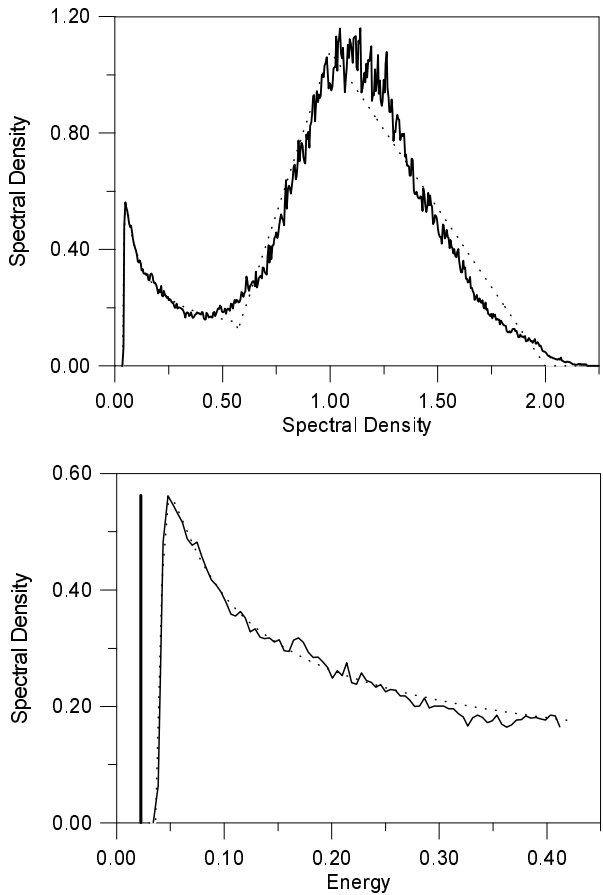}
\caption{Model spectral density (dashed line) and the result of spectral 
analysis (solid line). The position of the delta-function is shown only
in the lower panel.}
\label{fig:ap_sp_1}
\end{figure}

To check the accuracy of our approach, we tested it for the spectral 
density distribution that spreads over large range of frequencies and 
simultaneously possesses fine structure in low-frequency region. 
The test spectrum was modeled as the sum of the delta-function with the 
energy $\varepsilon_{\delta}=0.03$ and the weight $Z_{\delta}=0.07$, and 
continuous high-frequency spectral density which starts at the threshold
$\varepsilon_{\mbox{\scriptsize th}}=0.04$. The continuous part of the 
spectrum $\rho_{\mbox{\scriptsize con}}$ was modeled by the function
[In fact, this functional form
is predicted by the Pitaevskii theory for $\rho (\omega )$ near the end point.]
\begin{equation}
\rho_{\mbox{\scriptsize con}}(\omega) = 
\frac{Z_{\delta}\sqrt{\omega-\varepsilon_{\mbox{\scriptsize th}}}}
{2\pi\sqrt{\varepsilon_{\mbox{\scriptsize gap}}}
[(\omega-\varepsilon_{\mbox{\scriptsize th}})
+\varepsilon_{\mbox{\scriptsize gap}}]}
\label{ap_sp24}
\end{equation}
(here 
$\varepsilon_{\mbox{\scriptsize gap}} = 
\varepsilon_{\mbox{\scriptsize th}} - \varepsilon_{\delta}$
is a microgap)    
in the range $\omega \in [\varepsilon_{\mbox{\scriptsize th}}, 0.566]$ 
and by a triangule at higher frequencies (see the dashed line in the 
upper panel of Fig.~\ref{fig:ap_sp_1}). 
The Green function $G(\tau)$ was calculated 
from the model spectral density in the $n_{\mbox{\scriptsize max}}=300$ 
points 
$\tau_i = \tau_{\mbox{\scriptsize max}} i^2 / n_{\mbox{\scriptsize max}}$ 
in the time range from zero to $\tau_{\mbox{\scriptsize max}}=1000$. 
The restored spectral density reproduces 
both gross features of high-frequency part (upper panel in 
Fig.~\ref{fig:ap_sp_1})
and the fine structure at small frequencies (lower panel of 
Fig.~\ref{fig:ap_sp_1}). 
The energy and the weight of the delta-function was restored with the 
accuracy $10^{-4}$. The final solution was obtained by the averaging 
(\ref{ap_sp4}) of $M=1100$ particular solutions. 
\begin{figure}
\epsfxsize=0.95\textwidth
\epsfbox{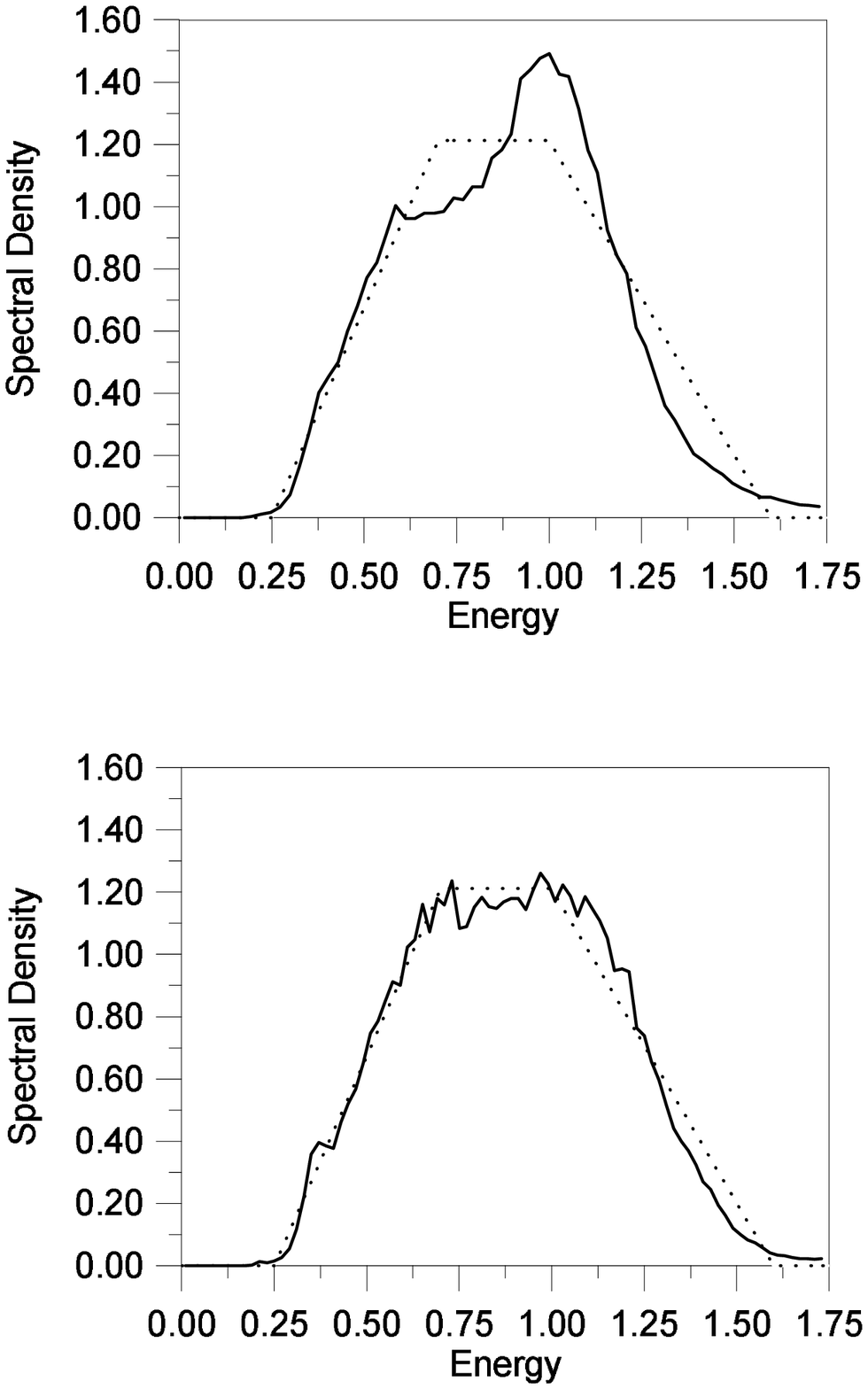}
\caption{The model spectrum (dashed lines) and results of spectral analysis
(solid lines) for $\eta=10^{-2}$ (upper panel) and $\eta=10^{-3}$ (lower
panel).}
\label{fig:ap_sp_2}
\end{figure}

To evaluate the precision which characterizes how typical particular 
solution $\tilde{\rho}_{j}(\omega )$ (see (\ref{ap_sp3})) reproduces the Green 
function $G(\tau)$, we introduce the maximal relative deviation 
\begin{equation}
\eta = \max \left[
\frac{\mid G(\tau_i) - \tilde{G}(\tau_i) \mid}
{G(\tau_i)}
\right], \; i \in [1,n_{\mbox{\scriptsize max}}] \; , 
\label{ap_sp25}
\end{equation}
which typical value is $\eta=10^{-4}$ for a particular solution of the 
spectrum in Fig.~\ref{fig:ap_sp_1}.

Since the Green function which is obtained from Monte Carlo calculations 
contains some statistic errors at each time point, the minimal value of 
parameter $\eta$ is limited by the quality of the calculated Green function 
$G(\tau)$. To study the influence of the (uncorrelated) statistic errors
we studied the stability of the method against stochastic noise 
\begin{equation}
G(\tau_i) \to G(\tau_i) (1+\eta r_i) \;, \; 
i=1,n_{\mbox{\scriptsize max}} \; ,
\label{ap_sp26}
\end{equation}
introduced by 
random numbers $r_i \in [0,1]$. It is seen that the method 
restores the gross features of the spectrum (position and width)
even for rather roughly calculated Green function, with $\eta=10^{-2}$ (upper 
panel in Fig.~\ref{fig:ap_sp_2}), whereas the precision $\eta=10^{-3}$ is 
sufficient to resolve the lineshape (lower panel in Fig.~\ref{fig:ap_sp_2})).

\newpage
 

\end{document}